\documentclass{article}%
\usepackage{epsfig}
\usepackage{amsmath}
\usepackage{graphicx}%
\usepackage{amsfonts}%
\usepackage{amssymb}
\begin{document}

\date{}%

\begin{titlepage}
\begin{center}
{\LARGE\bf3-Body Dynamics in a (1+1) Dimensional Relativistic Self-Gravitating
System} \\ \vspace{2cm} J.J. Malecki\footnotemark\footnotetext{email:
jjmaleck@uwaterloo.ca} and R.B. Mann\footnotemark\footnotetext{email:
mann@avatar.uwaterloo.ca} \\ \vspace{0.5cm} Dept. of Physics, University of
Waterloo Waterloo, ONT N2L 3G1, Canada\\ \vspace{2cm} PACS numbers: 04.40.-b,
04.25.-g, 05.45.Ac, 04.20.Jb\\ \vspace{2cm} \today\\
\end{center}
\begin{abstract}
The results of our study of the motion of a three particle, self-gravitating
system in general relativistic lineal gravity is presented for an arbitrary
ratio of the particle masses.  We derive a canonical expression for the
Hamiltonian of the system and discuss the numerical solution of the resulting
equations of motion.  This solution is compared to the corresponding
non-relativistic and post-Newtonian approximation solutions so that the
dynamics of the fully relativistic system can be interpretted as a correction
to the one-dimensional Newtonian self-gravitating system.  We find that the
structure of the phase space of each of these systems yields a large variety
of interesting dynamics that can be divided into three distinct regions:
annulus, pretzel, and chaotic; the first two being regions of
quasi-periodicity while the latter is a region of chaos. By changing the
relative masses of the three particles we find that the relative sizes of
these three phase space regions changes and that this deformation can be
interpreted physically in terms of the gravitational interactions of the
particles.  Furthermore, we find that many of the interesting characteristics
found in the case where all of the particles share the same mass also appears
in our more general study. We find that there are additional regions of chaos
in the unequal mass system which are not present in the equal mass case. We
compare these results to those found in similar systems.
\end{abstract}
\end{titlepage}

\section{Introduction}

\label{sec:intro} The calculation of the motion of $N$ particles under a
specified, mutual force is one of the oldest problems in physics, commonly
referred to as the $N$-body problem. This problem occurs frequently in many
distinct subfields and remains an active area of research. When the specified
force is that of Newtonian gravity in three spatial dimensions, a closed form
solution of the motion can be obtained for $N=2$. This is not true for (3+1)
dimensional general relativistic gravity, however, due to the existence of
energy dissipation in the form of gravitational radiation. All attempts to
calculate the motion of more than one particle in (3+1) general relativity
have required some form of approximation.

Considerable progress in this area of research has been made recently by
reducing the number of spatial dimensions from three to one. These
lower-dimensional theories provide a simpler prototype for their
higher-dimensional counterparts. Furthermore, for Newtonian gravity,
one-dimensional self-gravitating systems (OGS's) have proven to be very useful
in modelling many diverse physical systems. For example, it has been found
that there exist stable core-halo structures in the OGS phase-space that are
reminiscent of those found in globular clusters \cite{yawn}. These structures
consist of a dense core of particles near equilibrium surrounded by a cloud of
high kinetic energy particles that interact very weakly with the core. The OGS
also models the dynamics of flat, parallel sheets colliding along a
perpendicular axis \cite{LMiller} and the motion of stars interacting with a
highly flattened galaxy \cite{Rybicki}. More specifically, the three-particle
OGS has been found to model the motion of a billiard colliding with a wedge in
a uniform, gravitational field \cite{LMiller}, two elastically colliding
billiard balls in a uniform, gravitational field \cite{Goodings}, and a bound
state of three quarks to form a ``linear baryon'' \cite{Bukta}. There are
still many open questions about the OGS concerning its ergodic behaviour, the
conditions (if any) under which equipartition of energy is attained, whether
or not it can reach a true equilibrium configuration from arbitrary initial
conditions, and the appearance of fractal behaviour \cite{fractal}.

In a relativistic context, reduction of the number of spatial dimensions
results in an absence of gravitational radiation while retaining most (if not
all) of the remaining conceptual features of relativistic gravity.
Consequently, one might hope to obtain insight into the nature of relativistic
dynamical gravitational systems at the classical (and perhaps even quantum)
level in a wide variety of physical situations by studying the relativistic
OGS, or ROGS.

Comparatively little has been known about the ROGS (even for $N=2$) until
quite recently, when a prescription for obtaining its Hamiltonian from a
generally covariant, minimally-coupled action was obtained \cite{OR}. In the
non-relativistic limit ($c\rightarrow\infty$), the Hamiltonian reduces to that
of the OGS. This opened up the possibility of extending the insights of the
OGS into the relativistic regime, and indeed, considerable progress has been
made. Exact, closed-form solutions to the $2$-body problem have been obtained
\cite{2bd}. These have been extended to include both a cosmological constant
\cite{2bdcossh,2bdcoslo}\ and electromagnetic interactions \cite{2bdchglo},
and a new exact solution to the static-balance problem has been obtained
\cite{statbal}. In the $N$-body case, the Hamiltonian can be obtained as a
series expansion in inverse powers of the speed of light $c$ to arbitrary
order and a complete derivation of the partition and single-particle
distribution functions has been found in both the canonical and microcanonical
ensembles \cite{pchak} providing interesting information concerning the
influence of relativistic effects on self-gravitating systems. Very recently,
formulation of the ROGS has been extended to circular topologies \cite{circle}
(forbidden for the OGS), and a new $N$-body dynamical equilibrium solution has
been found \cite{ryan}. An exact expression for the relativistic, 3-body
Hamiltonian has been calculated and the motion of three equal-mass particles
has been extensively studied \cite{3bdshort,burnell}; these results are
summarized in \S~\ref{sec:eqmassoln}. \newline

\bigskip

In this paper, we will generalize the study of the motion of three particles
to include the unequal mass case. We work with a 2D theory of gravity on a
line (lineal gravity) that models 4D general relativity in that it sets the
Ricci scalar $R$ equal to the trace of the stress-energy of prescribed matter
fields and sources. Hence, as in $(3+1)$ dimensions, the evolution of
spacetime curvature is governed by the matter distribution, which in turn is
governed by the dynamics of spacetime \cite{r3}. Sometimes referred to as
$R=T$ theory, it is a particular member of a broad class of dilaton gravity
theories formulated on a line. What singles it out for consideration is its
consistent non-relativistic (i.e. $c\rightarrow\infty$ ) limit \cite{r3} which
is, in general, a problematic limit for a generic $(1+1)$-dimensional theory
of gravity \cite{jchan}. Consequently it contains each of the aforementioned
non-relativistic self-gravitating systems as special cases. Furthermore, it
reduces to Jackiw-Teitelboim (JT) theory \cite{JT} when the stress-energy is
that of a cosmological constant.

We have found that the best way to study the motion of three particles is to
work in the canonical formalism. By expressing the action in canonical
variables we are able to determine the Hamiltonian as a spatial integral of
the second derivative of the dilaton field. This field is determined by the
constraint equations derived from the action which can be solved by matching
the solution of the field across each of the three particles. The result is a
transcendental equation containing the Hamiltonian and expressed in terms of
the remaining degrees of freedom, that is, the two mutual separations of the
particles and their conjugate momenta. From this transcendental equation we
obtain the canonical equations of motion, which are then solved numerically.

Through a change of coordinates, the Newtonian, three-particle OGS can be
shown to be isomorphic to the motion of a single particle in a linear,
hexagonal well potential. By applying this same change of variables to the
three-particle ROGS we find an analogous hexagonal potential where the sides
of the hexagonal cross-section are curved outwards and the sides of the well
no longer increase linearly with increasing particle separation. We find that,
by changing the relative masses of the particles, the shape of the hexagonal
cross-section in both the Newtonian and relativistic systems is expanded or
contracted perpendicular to one of the lines connecting opposite vertices.
This change of variables simplifies the analysis of the motion significantly
and is used throughout to extract useful information from both the
three-particle OGS and ROGS.

As in ref.~\cite{burnell} we consider three distinct 3-body, self-gravitating
systems: the non-relativistic case (N) which has been extensively studied in
many different contexts \cite{LMiller, Goodings, Bukta, MillRavi}%
\footnote{These studies examine the 3-body problem in a classical potential
obtained by solving Poisson's equation in one spatial dimension. This
potential linearly depends on the separation of the particles as seen
in~(\ref{Hnewt}). The chaotic properties of the one dimensional 3-body problem
with a potential that depends inversely on the separation (as in three
dimensions) have been studied in~\cite{1overR}.}, the fully relativistic case
(R) described above, and a post-Newtonian expansion (pN) of the R system,
truncated to leading order in $c^{-2}$. While exact relativistic solutions of
the N-body problem have only been found for $N=2,3$, the post-Newtonian
expansion has been found for all finite values of $N$ up to any order of
accuracy \cite{OR}. Both the R and pN systems reduce to the N system in the
limit $c\rightarrow\infty$.

In section~\ref{sec:hamform} we outline the canonical reduction procedure of
\cite{burnell} that leads to the relativistic Hamiltonian expression and the
resulting canonical equations of motion. Some general properties of each of
the systems are then discussed in section~\ref{sec:genprop}, focussing on the
character of the associated potential energy functions of each. The method for
numerically solving the equations of motion is described in
section~\ref{sec:methsolv} and the results of this numerical solution
presented in section~\ref{sec:solvequ}. These results are then summarized and
discussed in section~\ref{sec:disc}, which concludes with a comment on areas
of further research interest.

\section{Hamiltonian Formulation of the Relativistic Equations of Motion}

\label{sec:hamform} The general procedure for deriving the $N$-body
Hamiltonian via canonical reduction is given in \cite{2bd, 2bdcoslo, burnell}
and only a brief description will be given here.

The action for the gravitational field minimally coupled to $N$ point
particles in (1+1) dimensions is given by
\begin{align}
I  &  =\int d^{2}x\left[  \frac{1}{2\kappa}\sqrt{-g}g^{\mu\nu}\left\{  \Psi
R_{\mu\nu}+\frac{1}{2}\nabla_{\mu}\Psi\nabla_{\nu}\Psi\right\}  \right.
\label{act1}\\
&  \makebox[2em]{}\left.  +\sum_{a=1}^{N}\int d\tau_{a}\left\{  -m_{a}\left(
-g_{\mu\nu}(x)\frac{dz_{a}^{\mu}}{d\tau_{a}}\frac{dz_{a}^{\nu}}{d\tau_{a}%
}\right)  ^{2}\right\}  \delta^{\left(  2\right)  }(x-z_{a}(\tau_{a}))\right]
\;,\nonumber
\end{align}
where $g_{\mu\nu}$ is the metric tensor with determinant $g$, $R_{\mu\nu} $ is
the Ricci tensor, $\tau_{a}$ the proper time for the $a$th particle with mass
$m_{a}$ and position $z_{a}$, and $\kappa=8\pi G/c^{4}$. We use $\nabla_{\mu}$
to denote the covariant derivative associated with $g_{\mu\nu}$. The scalar
(dilaton) field $\Psi$ has been incorporated because the classical
Einstein-Hilbert action in (1+1) dimensions is trivial due to the vanishing of
the Einstein tensor. This action describes a self-gravitating system of $N$
particles without collisional terms (\textit{i.e.} the particles pass through
each other).

>From (\ref{act1}) one can derive the following field equations
\begin{align}
&  R-g^{\mu\nu}\nabla_{\mu}\nabla_{\nu}\Psi=0\;,\label{eq-R}\\
&  \frac{1}{2}\nabla_{\mu}\Psi\nabla_{\nu}\Psi-\frac{1}{4}g_{\mu\nu}%
\nabla^{\lambda}\Psi\nabla_{\lambda}\Psi+g_{\mu\nu}\nabla^{\lambda}%
\nabla_{\lambda}\Psi-\nabla_{\mu}\nabla_{\nu}\Psi=\kappa T_{\mu\nu
},\label{Psieq}\\
&  m_{a}\left[  \frac{d}{d\tau_{a}}\left\{  g_{\mu\nu}(z_{a})\frac{dz_{a}%
^{\nu}}{d\tau_{a}}\right\}  -\frac{1}{2}g_{\nu\lambda,\mu}(z_{a})\frac
{dz_{a}^{\nu}}{d\tau_{a}}\frac{dz_{a}^{\lambda}}{d\tau_{a}}\right]  =0\;,
\label{eq-z}%
\end{align}
where
\begin{equation}
T_{\mu\nu}=\sum_{a}m_{a}\int d\tau_{a}\frac{1}{\sqrt{-g}}g_{\mu\sigma}%
g_{\nu\rho}\frac{dz_{a}^{\sigma}}{d\tau_{a}}\frac{dz_{a}^{\rho}}{d\tau_{a}%
}\delta^{\left(  2\right)  }(x-z_{a}(\tau_{a}))\; \label{stressenergy}%
\end{equation}
is the stress-energy tensor for the $N$ particles and is conserved via
(\ref{Psieq}). Inserting the trace of (\ref{Psieq}) into (\ref{eq-R}) we
obtain
\begin{equation}
R=\kappa T_{\;\;\mu}^{\mu}. \label{RT}%
\end{equation}
The fact that we retain this simple relation between the geometry of spacetime
and the matter, analogous to the Einstein field equations, is the motivation
for choosing the dilaton coupling in (\ref{act1}).

Equations (\ref{eq-z}) and (\ref{RT}) form a $N+1$ system that can be solved
for the single metric degree of freedom and the $N$ particle degrees of
freedom. Equation (\ref{Psieq}) relates the evolution of the dilaton field to
the evolution of the point masses.

\bigskip To arrive at a Hamiltonian theory we begin by writing the metric as
\begin{equation}
ds^{2}=-N_{0}^{2}\left(  x,t\right)  dt^{2}+\gamma\left(  dx+\frac{N_{1}%
}{\gamma}dt\right)  ^{2} \label{metric}%
\end{equation}
where $N_{0}$ and $N_{1}$ are the lapse and shift functions which act as
Lagrange multipliers for the resulting constraints of the Hamiltonian system
and $\gamma$ is the single metric degree of freedom. \bigskip

By also defining $p_{a}$, $\pi$ and $\Pi$ to be the conjugate momentum of
$z_{a}$, $\gamma$ and $\Psi$ respectively, one can canonically reduce the
action to the form \cite{OR}
\begin{equation}
I=\int d^{2}x\left\{  \sum_{a}p_{a}\dot{z}_{a}\delta(x-z_{a})+\frac{1}{\kappa
}\triangle\Psi\right\}  \; \label{actred}%
\end{equation}
upon eliminating the constraints and choosing the coordinate conditions
$\gamma=1$ and $\Pi=0$. Here we use $\triangle$ to denote $\partial
^{2}/\partial x^{2}$ and a dot to denote $\partial/\partial t$. With the
action in this form, we recognize the second term $\mathcal{H}=-\frac
{1}{\kappa}\triangle\Psi$ as the Hamiltonian density and can immediately write
down the Hamiltonian for $N$ particles as
\begin{equation}
H=\int dx\mathcal{H}=-\frac{1}{\kappa}\int dx\triangle\Psi\; \label{ham1}%
\end{equation}
where $\Psi$ is a function of $z_{a}$ and $p_{a}$ and can be determined from
the solution to the constraint equations which now take the form
\begin{equation}
\triangle\Psi-\frac{1}{4}(\Psi^{\prime})^{2}+\kappa^{2}\pi^{2}+\kappa\sum
_{a}\sqrt{p_{a}^{2}+m_{a}^{2}}\delta(x-z_{a})=0 \label{cst1}%
\end{equation}%
\begin{equation}
2\pi^{\prime}+\sum_{a}p_{a}\delta(x-z_{a})=0\;\; \label{cst2}%
\end{equation}
where a prime denotes $\partial/\partial x$.

The solution of (\ref{ham1}), (\ref{cst1}), and (\ref{cst2}) for the 3
particle case is given in \cite{burnell} and will not be reproduced here in
detail. The basic procedure involves choosing a specific configuration of the
3 particles and solving (\ref{cst1}) and (\ref{cst2}) in the region between
each particle. The constants of integration are then determined by demanding
that $\Psi$ and $\Psi^{\prime}$ remain finite and coincide at the position of
the particles. This gives an implicit equation for the Hamiltonian $H$ for the
specified particle configuration and the Hamiltonian for a general
configuration is obtained by permutation of the particle indices (1, 2, and 3).

This implicit equation for the Hamiltonian can be expressed as
\begin{align}
L_{1}L_{2}L_{3}  &  =\mathcal{M}_{12}\mathcal{M}_{21}L_{3}^{\ast}%
e^{\frac{\kappa}{4}s_{12}[(L_{1}+\mathcal{M}_{12})z_{13}-(L_{2}+\mathcal{M}%
_{21})z_{23}]}\nonumber\\
&  +\mathcal{M}_{23}\mathcal{M}_{32}L_{1}^{\ast}e^{\frac{\kappa}{4}%
s_{23}[(L_{2}+\mathcal{M}_{23})z_{21}-(L_{3}+\mathcal{M}_{32})z_{31}%
]}\nonumber\\
&  +\mathcal{M}_{31}\mathcal{M}_{13}L_{2}^{\ast}e^{\frac{\kappa}{4}%
s_{31}[(L_{3}+\mathcal{M}_{31})z_{32}-(L_{1}+\mathcal{M}_{13})z_{12}]}
\label{Htrans}%
\end{align}
or more compactly
\begin{equation}
L_{1}L_{2}L_{3}= \frac{1}{2} \sum_{ijk}\left|  \epsilon^{ijk}\right|
\mathcal{M}_{ij}\mathcal{M}_{ji}L_{k}^{\ast}e^{\frac{\kappa}{4}s_{ij}%
[(L_{i}+\mathcal{M}_{ij})z_{ik}-(L_{j}+\mathcal{M}_{ji})z_{jk}]}
\label{Htranscomp}%
\end{equation}
where
\begin{align}
\mathcal{M}_{ij}  &  =M_{i}-\epsilon p_{i}s_{ij}%
,{\ \ \ \ \ \ \ \ \ \ \ \ \ \ \ \ \ \ \ \ \ \ \ \ }M_{i}=\sqrt{p_{i}^{2}%
+m_{i}^{2}}\label{massM}\\
L_{i}  &  =H-M_{i}-\epsilon(\sum_{j}p_{j}s_{ji}){\ \ \ \ \ \ \ \ \ }%
L_{i}^{\ast}=(1-\prod_{j<k\neq i}s_{ij}s_{ik})M_{i}+L_{i} \label{Ldef}%
\end{align}
with $z_{ij}=(z_{i}-z_{j})$ , $s_{ij}={sgn}(z_{ij})$, and $\epsilon^{ijk}$ is
the 3-dimensional Levi-Civita tensor. The discrete parameter $\epsilon=\pm1$
is a constant of integration that flips sign under time reversal. This
provides a measure of the flow of time of the gravitational field relative to
the particle momenta.

Although we cannot obtain an explicit expression for the Hamiltonian, we are
able to derive the equations of motion explicitly by partially differentiating
(\ref{Htranscomp}) implicitly with respect to $z_{a}$ and $p_{a}$ and solving
for $\partial H / \partial z_{a}$ and $\partial H / \partial p_{a}$
respectively. From Hamilton's equations
\begin{align}
\dot{z}_{a}  &  =\frac{\partial H}{\partial p{_{a}}}\label{zdot}\\
\dot{p}_{a}  &  =-\frac{\partial H}{\partial z{_{a}}} \label{pdot}%
\end{align}
we can obtain the equations of motion.

For example, for $a = 1$, (\ref{zdot}) and (\ref{pdot}) become
\begin{align}
&  \dot{z}_{1}\left\{  \phantom{\frac{}{}} L_{2}L_{3}+L_{1}L_{3}+L_{1}%
L_{2}\right. \nonumber\\
&  -[M_{2}-\epsilon p_{2}s_{21}][M_{1}-\epsilon p_{1}s_{12}][1+\frac{\kappa
}{4}L_{3}^{\ast}|z_{12}|]e^{\frac{\kappa}{4}s_{12}[(L_{1}+\mathcal{M}%
_{12})z_{13}-(L_{2}+\mathcal{M}_{21})z_{23}]}\nonumber\\
&  -[M_{3}-\epsilon p_{3}s_{31}][M_{1}-\epsilon p_{1}s_{13}][1+\frac{\kappa
}{4}L_{2}^{\ast}|z_{13}|]e^{\frac{\kappa}{4}s_{13}[(L_{1}+\mathcal{M}%
_{23})z_{12}+(L_{3}+\mathcal{M}_{32})z_{23}]}\nonumber\\
&  \left.  -[M_{2}-\epsilon p_{2}s_{23}][M_{3}-\epsilon p_{3}s_{32}%
][1+\frac{\kappa}{4}L_{1}^{\ast}|z_{23}|]e^{\frac{\kappa}{4}s_{23}%
[(L_{3}+\mathcal{M}_{31})z_{13}-(L_{2}+\mathcal{M}_{13})z_{12}]}\right\}
\nonumber\\
&  =[M_{2}-\epsilon p_{2}s_{21}][(\frac{\partial M_{1}}{\partial p_{1}%
}-\epsilon s_{12})L_{3}^{\ast}-(M_{1}-\epsilon p_{1}s_{12})(\epsilon
s_{13}+\frac{\kappa}{4}L_{3}^{\ast}(\epsilon z_{12}))]e^{\frac{\kappa}%
{4}s_{12}[(L_{1}+\mathcal{M}_{12})z_{13}-(L_{2}+\mathcal{M}_{21})z_{23}%
]}\nonumber\\
&  +[M_{3}-\epsilon p_{3}s_{31}][(\frac{\partial M_{1}}{\partial p_{1}%
}-\epsilon s_{13})L_{2}^{\ast}-(M_{1}-\epsilon p_{1}s_{13})\{\epsilon
s_{12}+\frac{\kappa}{4}L_{2}^{\ast}(\epsilon z_{13})\}]e^{\frac{\kappa}%
{4}s_{13}[(L_{1}+\mathcal{M}_{23})z_{12}+(L_{3}+\mathcal{M}_{32})z_{23}%
]}\nonumber\\
&  +[M_{2}-\epsilon p_{2}s_{23}][M_{3}-\epsilon p_{3}s_{32}][-s_{12}%
s_{13}\frac{\partial M_{1}}{\partial p_{1}}+\frac{\kappa}{4}s_{23}L_{1}^{\ast
}[\epsilon|z_{12}|-\epsilon|z_{13}|]]e^{\frac{\kappa}{4}s_{23}[(L_{3}%
+\mathcal{M}_{31})z_{13}-(L_{2}+\mathcal{M}_{13})z_{12}]}\nonumber\\
&  +\frac{\partial M_{1}}{\partial p_{1}}L_{2}L_{3}+\epsilon(s_{12}L_{1}%
L_{3}+s_{13}L_{2}L_{1}) \label{z1dot}%
\end{align}
and
\begin{align}
&  \dot{p}_{1}\left\{  \phantom{\frac{}{}}L_{2}L_{3}+L_{1}L_{3}+L_{1}%
L_{2}\right. \nonumber\\
&  -[M_{2}-\epsilon p_{2}s_{21}][M_{1}-\epsilon p_{1}s_{12}][1+\frac{\kappa
}{4}L_{3}^{\ast}|z_{12}|]e^{\frac{\kappa}{4}s_{12}[(L_{1}+\mathcal{M}%
_{12})z_{13}-(L_{2}+\mathcal{M}_{21})z_{23}]}\nonumber\\
&  -[M_{3}-\epsilon p_{3}s_{31}][M_{1}-\epsilon p_{1}s_{13}][1+\frac{\kappa
}{4}L_{2}^{\ast}|z_{13}|]e^{\frac{\kappa}{4}s_{13}[(L_{1}+\mathcal{M}%
_{23})z_{12}+(L_{3}+\mathcal{M}_{32})z_{23}]}\nonumber\\
&  \left.  -[M_{2}-\epsilon p_{2}s_{23}][M_{3}-\epsilon p_{3}s_{32}%
][1+\frac{\kappa}{4}L_{1}^{\ast}|z_{23}|]e^{\kappa/4s_{23}[(L_{3}%
+M_{3}-\epsilon p_{3}s_{32})z_{13}-(L_{2}+M_{2}-\epsilon p_{2}s_{23})z_{12}%
]}\right\} \nonumber\\
&  =[M_{2}-\epsilon p_{2}s_{21}][M_{1}-\epsilon p_{1}s_{12}][\frac{\kappa}%
{4}s_{12}L_{3}^{\ast}[H+\epsilon(p_{2}-p_{1})s_{12}+\epsilon p_{3}%
s_{13}]]e^{\frac{\kappa}{4}s_{12}[(L_{1}+\mathcal{M}_{12})z_{13}%
-(L_{2}+\mathcal{M}_{21})z_{23}]}\nonumber\\
&  +[M_{3}-\epsilon p_{3}s_{31}][M_{1}-\epsilon p_{1}s_{13}][\frac{\kappa}%
{4}s_{13}L_{2}^{\ast}[H+\epsilon p_{2}s_{12}+\epsilon(p_{3}-p_{1}%
)s_{13}]]e^{\frac{\kappa}{4}s_{13}[(L_{1}+\mathcal{M}_{23})z_{12}%
+(L_{3}+\mathcal{M}_{32})z_{23}]}\nonumber\\
&  +[M_{2}-\epsilon p_{2}s_{23}][M_{3}-\epsilon p_{3}s_{32}][\frac{\kappa}%
{4}s_{23}L_{1}^{\ast}p_{1}(s_{12}-s_{13})]e^{\frac{\kappa}{4}s_{23}%
[(L_{3}+\mathcal{M}_{31})z_{13}-(L_{2}+\mathcal{M}_{13})z_{12}]}.
\label{pdot1}%
\end{align}
The equations for $a = 2, 3$ are similar and will be omitted here.

\section{General Properties of the 3-Body System}

\label{sec:genprop} Before we go on to solve the equations of motion, it is
instructive to consider some general characteristics of the 3-body system
described by the determining equation (\ref{Htrans}) and its associated
non-relativistic and post-Newtonian counterparts.

To compare the relativistic motion to that predicted classically, we introduce
the Newtonian $N$ particle Hamiltonian in (1+1) dimensions
\begin{equation}
H_{N} = \sum_{a} \frac{p_{a}^{2}}{2m_{a}} + \pi G \sum_{a} \sum_{b} m_{a}
m_{b} \left|  z_{ab} \right|  . \label{Hnewt}%
\end{equation}
where $z_{ab} = z_{a} - z_{b}$ as before. We shall refer to this as the
Newtonian or N system.

A post-Newtonian approximation of the general $N$-body Hamiltonian has been
found \cite{OR} and is given here up to order $c^{-2}$
\begin{align}
H_{pN}  &  = c^{2} \sum_{a} m_{a} + \sum_{a} \frac{p_{a}^{2}}{2m_{a}} +
\frac{\kappa c^{4}}{8} \sum_{a} \sum_{b} m_{a} m_{b} \left|  z_{ab} \right|  +
\frac{\epsilon\kappa c^{3}}{8} \sum_{a} \sum_{b} \left(  m_{a} p_{b} - m_{b}
p_{a}\right)  \left(  z_{ab}\right)  - {}\nonumber\\
&  - c \sum_{a} \frac{p_{a}^{4}}{8 m_{a}^{3}} + \frac{\kappa c^{2}}{8}
\sum_{a} \sum_{b} \left(  m_{a} \frac{p_{a}^{2}}{m_{b}} \left|  z_{ab}
\right|  - p_{a} p_{b} \left|  z_{ab} \right|  \right)  + {}\nonumber\\
&  + \frac{1}{4} \left(  \frac{\kappa}{4}\right)  ^{2} c^{6} \sum_{a} \sum_{b}
\sum_{c} m_{a} m_{b} m_{c} \left(  \left|  z_{ab}\right|  \left|
z_{ac}\right|  + z_{ab} z_{ac} \right)  . \label{upnham}%
\end{align}
If we re-scale (\ref{upnham}) to remove the constant $\sum_{a} m_{a} c^{2}$
term and take the limit $c \to\infty$ then it is clear that we retrieve the
Newtonian result (\ref{Hnewt}).

However, the coordinates $z_{a}$ and $p_{a}$ are not necessarily the most
natural coordinates to use to describe the post-Newtonian system. The reason
is that the fourth term on the right hand side of (\ref{upnham}) is
proportional to $c^{-1}$. In $(3+1)$ dimensions, terms in odd powers of
$c^{-1}$ are associated with gravitational radiation, but, in $(1+1)$
dimensions, there are not enough degrees of freedom to allow for the existence
of gravitational radiation, and such terms are artifacts of the choice of
canonical variables.

Indeed, as in \cite{OR}, we can remove the $c^{-1}$ term by performing the
canonical transformation
\begin{equation}
z_{a}\rightarrow\tilde{z}_{a}=z_{a} \label{zatilde}%
\end{equation}%
\begin{equation}
p_{a}\rightarrow\tilde{p}_{a}=p_{a}-\frac{\epsilon\kappa}{4}\sum_{b}m_{a}%
m_{b}z_{ab} \label{patilde}%
\end{equation}
after which, (\ref{upnham}) becomes
\begin{align}
\tilde{H}_{pN}  &  =c^{2}\sum_{a}m_{a}+\sum_{a}\frac{\tilde{p}_{a}^{2}}%
{2m_{a}}+\frac{\kappa c^{4}}{8}\sum_{a}\sum_{b}m_{a}m_{b}\left|  \tilde
{z}_{ab}\right|  -{}\nonumber\\
&  -c\sum_{a}\frac{\tilde{p}_{a}^{4}}{8m_{a}^{3}}+\frac{\kappa c^{2}}{8}%
\sum_{a}\sum_{b}\left(  m_{a}\frac{\tilde{p}_{a}^{2}}{m_{b}}\left|  \tilde
{z}_{ab}\right|  -\tilde{p}_{a}\tilde{p}_{b}\left|  \tilde{z}_{ab}\right|
\right)  +{}\nonumber\\
&  +\frac{1}{4}\left(  \frac{\kappa}{4}\right)  ^{2}c^{6}\sum_{a}\sum_{b}%
\sum_{c}m_{a}m_{b}m_{c}\left(  \left|  \tilde{z}_{ab}\right|  \left|
\tilde{z}_{ac}\right|  -\tilde{z}_{ab}\tilde{z}_{ac}\right)  \label{pnham}%
\end{align}
where $\tilde{z}_{ab}=\tilde{z}_{a}-\tilde{z}_{b}$. Since (\ref{pnham}) uses
different coordinates than (\ref{Hnewt}) and (\ref{Htrans}) it is important to
distinguish between the two expressions for the post-Newtonian Hamiltonian. We
will refer to the system described by (\ref{upnham}) as the untransformed
post-Newtonian system, or UpN system, and the system described by
(\ref{pnham}) simply as the post-Newtonian or pN system. Note that only the pN
system was studied in \cite{burnell}. To complete our nomenclature, the fully
relativistic system will be denoted as the R system. Unless otherwise stated,
the Newtonian system will be assumed to have been re-scaled so that
$H(z_{ab}=0,p_{a}=0)=(m_{1}+m_{2}+m_{3})c^{2}$ in order to properly compare it
to the pN and R cases.

In order to simplify our analysis we will adopt the convention of \cite{Bukta}
and \cite{burnell} and define the following canonical coordinates:
\begin{equation}
\rho=\frac{1}{\sqrt{2}}\left(  z_{1}-z_{2}\right)  ,{\ \ \ }\lambda=\frac
{1}{\sqrt{6}}\left(  z_{1}+z_{2}-2z_{3}\right)  ,{\ \ \ }Z=z_{1}+z_{2}+z_{3},
\label{rholambdef}%
\end{equation}
with conjugate momenta
\begin{equation}
p_{\rho}=\frac{1}{\sqrt{2}}\left(  p_{1}-p_{2}\right)  ,{\ \ \ }p_{\lambda
}=\frac{1}{\sqrt{6}}\left(  p_{1}+p_{2}-2p_{3}\right)  ,{\ \ \ }p_{Z}=\frac
{1}{3}\left(  p_{1}+p_{2}+p_{3}\right)  . \label{prholambdef}%
\end{equation}
In the non-relativistic limit, $Z$ and $p_{Z}$ are related to the center of
mass and its conjugate momentum. While the equivalence principle does not
allow us to arbitrarily set $Z$ in the relativistic case, we can, without loss
of generality, choose $p_{Z}$ to vanish. The consequence of this is that we
can explicitly express $p_{1}$, $p_{2}$, and $p_{3}$ in terms of the newly
defined momenta~(\ref{prholambdef}) but can only express the
\emph{separations} of the particles $z_{ab}$ explicitly in terms of the new
coordinates~(\ref{rholambdef}). This gives us the following relations:
\begin{equation}
z_{12}=\sqrt{2}\rho,{\ \ \ }z_{13}=\frac{1}{\sqrt{2}}(\sqrt{3}\lambda
+\rho),{\ \ \ }z_{23}=\frac{1}{\sqrt{2}}(\sqrt{3}\lambda-\rho),
\label{zijrholam}%
\end{equation}
\begin{equation}
p_{1}=\frac{1}{\sqrt{6}}p_{\lambda}+\frac{1}{\sqrt{2}}p_{\rho},{\ \ \ }%
p_{2}=\frac{1}{\sqrt{6}}p_{\lambda}-\frac{1}{\sqrt{2}}p_{\rho},{\ \ \ }%
p_{3}=-\sqrt{\frac{2}{3}}p_{\lambda}. \label{prholam}%
\end{equation}

All of the Hamiltonian expressions (\ref{Htrans}), (\ref{Hnewt}),
(\ref{upnham}), and (\ref{pnham}) do not depend on $Z$ or $p_{Z}$\ and so
these variables are irrelevant. Expressions for (\ref{Htrans}), (\ref{Hnewt}),
and (\ref{pnham}) in terms of the new coordinates are given in \cite{burnell}
for the case when $m_{1}=m_{2}=m_{3}$. The corresponding expressions for
unequal masses are very cumbersome and will not be reproduced here.

By defining the potential of each system as $V(\rho,\lambda)=H(p_{\rho
}=0,p_{\lambda}=0)$ we can compare some of the different characteristics of
the 3 systems. In $\rho$-$\lambda$ coordinates the N potential becomes
\begin{equation}
V_{N}=m_{1}+m_{2}+m_{3}+\frac{\kappa}{4\sqrt{2}}\left(  2m_{1}m_{2}\left|
\rho\right|  +m_{1}m_{3}\left|  \sqrt{3}\lambda+\rho\right|  +m_{2}%
m_{3}\left|  \sqrt{3}\lambda-\rho\right|  \right)  \label{npot}%
\end{equation}
where we have rescaled the Hamiltonian as described above (with $c$ henceforth
set to unity unless explicitly stated otherwise). The UpN potential is given
as
\begin{align}
V_{pN}  &  =m_{1}+m_{2}+m_{3}+\frac{\kappa}{4\sqrt{2}}\left(  2m_{1}%
m_{2}\left|  \rho\right|  +m_{1}m_{3}\left|  \sqrt{3}\lambda+\rho\right|
+m_{2}m_{3}\left|  \sqrt{3}\lambda-\rho\right|  \right) \nonumber\\
&  +\frac{1}{2}\left(  \frac{\kappa}{4}\right)  ^{2}m_{1}m_{2}m_{3}\left(
4\rho^{2}+\left(  \sqrt{3}\lambda+\rho\right)  ^{2}+\left(  \sqrt{3}%
\lambda-\rho\right)  ^{2}+\left(  1+s_{\rho}s_{1}\right)  \left|  \rho\right|
\left|  \sqrt{3}\lambda+\rho\right|  \right) \nonumber\\
&  +\frac{1}{2}\left(  \frac{\kappa}{4}\right)  ^{2}m_{1}m_{2}m_{3}\left(
\left(  1-s_{\rho}s_{2}\right)  \left|  \rho\right|  \left|  \sqrt{3}%
\lambda-\rho\right|  +\frac{1}{2}\left(  1+s_{1}s_{2}\right)  \left|  \sqrt
{3}\lambda+\rho\right|  \left|  \sqrt{3}\lambda-\rho\right|  \right)
\label{upnpot}%
\end{align}
and the pN potential is
\begin{align}
\tilde{V}_{pN}  &  =m_{1}+m_{2}+m_{3}+\frac{\kappa}{4\sqrt{2}}\left(
2m_{1}m_{2}\left|  \tilde{\rho}\right|  +m_{1}m_{3}\left|  \sqrt{3}%
\tilde{\lambda}+\tilde{\rho}\right|  +m_{2}m_{3}\left|  \sqrt{3}\tilde
{\lambda}-\tilde{\rho}\right|  \right) \nonumber\\
&  +\frac{1}{2}\left(  \frac{\kappa}{4}\right)  ^{2}m_{1}m_{2}m_{3}\left(
\left(  1-\tilde{s}_{\rho}\tilde{s}_{1}\right)  \left|  \tilde{\rho}\right|
\left|  \sqrt{3}\tilde{\lambda}+\tilde{\rho}\right|  +\left(  1+\tilde
{s}_{\rho}\tilde{s}_{2}\right)  \left|  \tilde{\rho}\right|  \left|  \sqrt
{3}\tilde{\lambda}-\tilde{\rho}\right|  \right. \nonumber\\
&  \left.  +\frac{1}{2}\left(  1-\tilde{s}_{1}\tilde{s}_{2}\right)  \left|
\sqrt{3}\tilde{\lambda}+\tilde{\rho}\right|  \left|  \sqrt{3}\tilde{\lambda
}-\tilde{\rho}\right|  \right)  \label{pnpot}%
\end{align}
where $\tilde{\rho}$ and $\tilde{\lambda}$ are defined as in (\ref{rholambdef}%
) using the $\tilde{z}_{a}$ coordinates of (\ref{zatilde}). Here we have
defined $s_{\rho}={sgn}(\rho)$, $s_{1}={sgn}\left(  \sqrt{3}\lambda
+\rho\right)  $, and $s_{2}={sgn}\left(  \sqrt{3}\lambda-\rho\right)  $ and
the $\tilde{s}$ terms are defined similarly in terms of $\tilde{\rho}$ and
$\tilde{\lambda}$. The exact relativistic potential can be calculated from
(\ref{Htrans}) to be
\begin{align}
\left(  V_{R}-m_{1}\right)  \left(  V_{R}-m_{2}\right)  \left(  V_{R}%
-m_{3}\right)   &  =m_{1}m_{2}\left(  V_{R}-s_{1}s_{2}m_{3}\right)
\exp\left[  \frac{\kappa}{2\sqrt{2}}V_{R}\left|  \rho\right|  \right]
\nonumber\\
&  +m_{1}m_{3}\left(  V_{R}+s_{\rho}s_{2}m_{2}\right)  \exp\left[
\frac{\kappa}{4\sqrt{2}}V_{R}\left|  \sqrt{3}\lambda+\rho\right|  \right]
\nonumber\\
&  +m_{2}m_{3}\left(  V_{R}-s_{\rho}s_{1}m_{1}\right)  \exp\left[
\frac{\kappa}{4\sqrt{2}}V_{R}\left|  \sqrt{3}\lambda-\rho\right|  \right]  .
\label{relpot}%
\end{align}

An extensive comparison between the different potentials has been given in
\cite{burnell} for the case where the particle masses are equal and so here we
wish to focus on the changes to the potential due to changes in the relative
masses of the 3 particles.

\begin{figure}[ptb]
\begin{center}
\epsfig{file=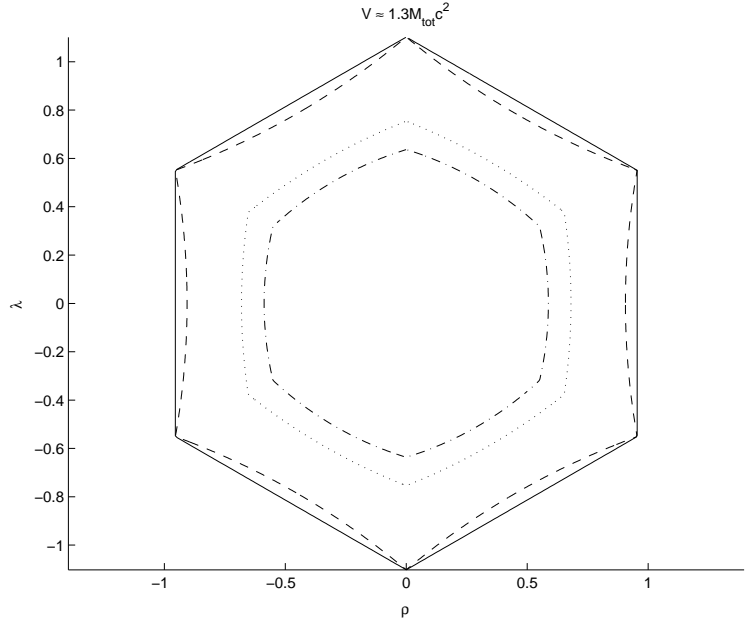, width=0.6\linewidth}
\end{center}
\caption{A cross section of the 4 potentials at $V\approx1.3M_{tot}c^{2}$ in
the case that all 3 particles have the same mass. N - solid, pN - dashed, UpN
- dotted, R - dash-dotted.  The $\rho$ and $\lambda$ are dimensionless variables defined using the  dimensionless positions $\hat{z}_i$ of equation~(\protect\ref{zhat}).}%
\label{fig:eqpot}%
\end{figure}

A cross section of each of the potentials at a fixed value of $V$ is shown in
Figure~\ref{fig:eqpot} for the case where all particles have the same mass.
All of the potentials share a certain hexagonal symmetry in that they are all
smooth except along the lines
\begin{align}
\rho &  =0\label{bisect1}\\
\rho+\sqrt{3}\lambda &  =0\label{bisect2}\\
\rho-\sqrt{3}\lambda &  =0 \label{bisect3}%
\end{align}
which correspond to $z_{1}=z_{2}$, $z_{1}=z_{3}$, and $z_{2}=z_{3}$
respectively (\textit{i.e.} the potential is not differentiable when two
particles are coincident). This is true for all ratios of the masses of the
particles. 

\begin{figure}[ptb]
\begin{center}
\epsfig{file=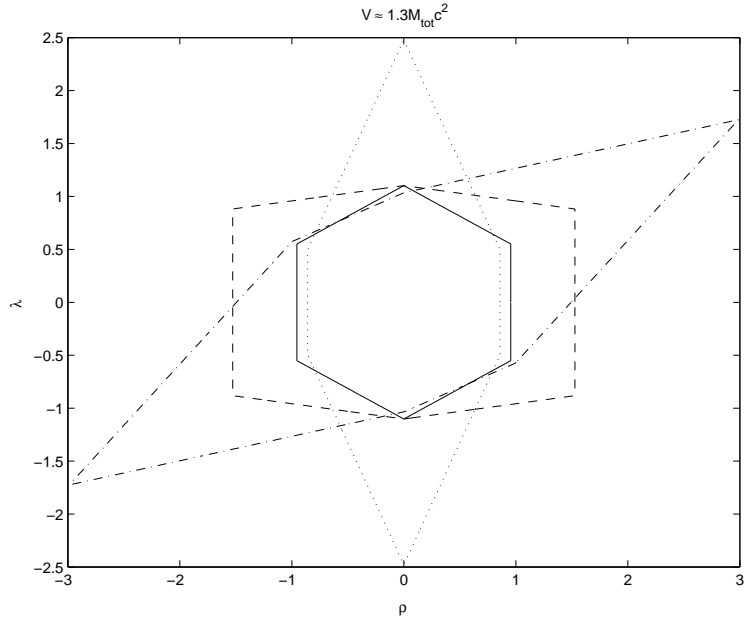, width=0.6\linewidth}
\end{center}
\caption{Cross-sections of the Newtonian potential at $V\approx1.3M_{{tot}%
}c^{2}$ for various mass ratios $m_{1}$:$m_{2}$:$m_{3}$. Solid - 1:1:1; dashed
- 1:1:4; dotted - 4:4:1; dash-dotted - 1:4:8. Notice that all discontinuities
lie on one of the three bisectors (\ref{bisect1}), (\ref{bisect2}), or
(\ref{bisect3}) regardless of the mass ratio.  The $\rho$ and $\lambda$ are dimensionless variables as in Figure~\protect\ref{fig:eqpot}.}%
\label{fig:newtpot}%
\end{figure}

The Newtonian potential is a distorted hexagonal well with sides that increase
linearly with $V$. The hexagonal cross-section at any value of $V_{N}$ only
has equal length sides when $m_{1}=m_{2}=m_{3}$. Figure~\ref{fig:newtpot}
shows various cross-sections of the Newtonian potential at a fixed value of
$V$ for different mass ratios. We see that increasing the mass of particle 3
has the effect of expanding the the hexagon away from the $\rho=0$ bisector
while decreasing the mass contracts the hexagonal cross-section towards
$\rho=0$. Increasing and decreasing the mass of particles 1 or 2 has the same
effect but the deformation is perpendicular to the $\rho-\sqrt{3}\lambda=0$ or
$\rho+\sqrt{3}\lambda=0$ respectively. When all three particles have unequal
mass, the hexagon is deformed as above with the magnitude of the deformation
in each direction given by the relative values of the mass.

\begin{figure}[ptb]
\begin{center}
\rotatebox{270}{\epsfig{file=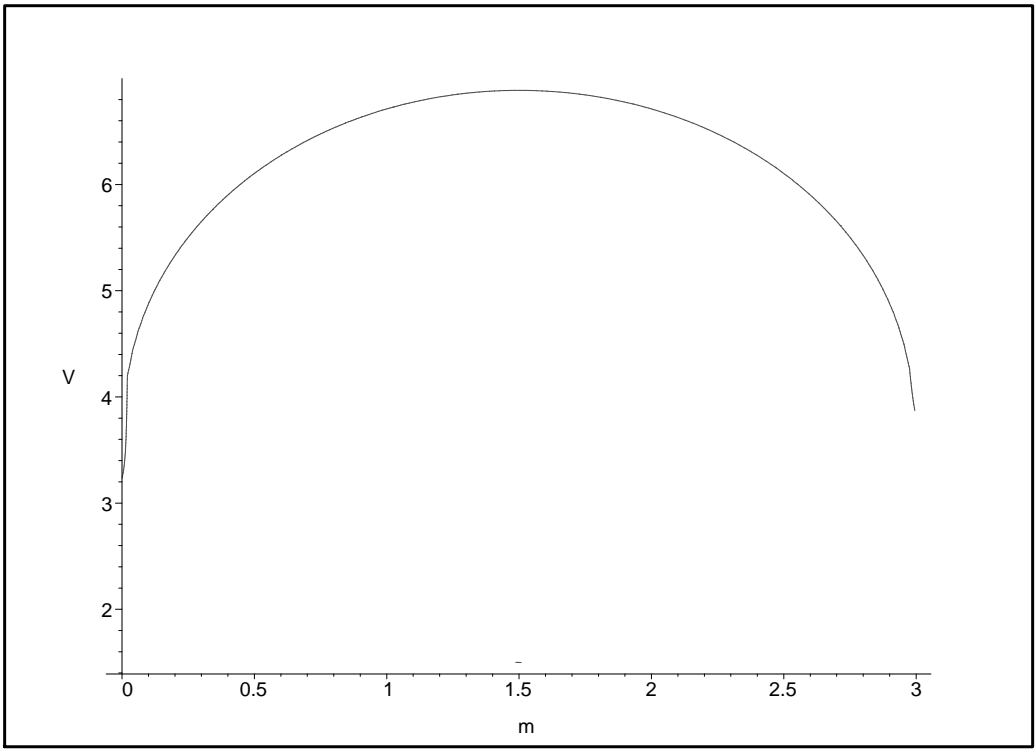, width=0.6\linewidth}}
\end{center}
\caption{Critical values of the relativistic potential $V_{R}$ as a function
of a given particle mass in units of $M_{\text{tot}}$ (here set equal to 3).
The maximum critical value occurs in the case when $m_{j} = M_{tot} / 2$.
\ The minimal value approaches the limit $V_{R}\simeq M_{\text{tot}}$ as
$m_{j}\rightarrow0$ or $M_{\text{tot}}$.}%
\label{fig:Vcritvsm}%
\end{figure}

The relativistic potential is similar to the Newtonian potential except that
the sides of the hexagon become concave. Furthermore, for small values of
$(\rho,\lambda)$, the relativistic potential increases much more rapidly than
the Newtonian potential. However, at a value $\hat{V}_{R}$ such that
\begin{equation}
\ln\left(  \frac{\left(  \hat{V}_{R}-m_{j}\right)  \left(  \hat{V}_{R}-\left(
M_{\text{tot} }-m_{j}\right)  \right)  }{\left(  M_{\text{tot}}-m_{j}\right)
m_{j}}\right)  =\hat{V}_{R}\left[  \frac{1}{\left(  \hat{V}_{R}-m_{j}\right)
}+\frac{1}{\left(  \hat{V}_{R}-\left(  M_{\text{tot}}-m_{j}\right)  \right)
}\right]  \label{Vcrit}%
\end{equation}
(for $j=1,2$ or $3$), the slope of the relativistic potential becomes
infinite, after which the size of the distorted hexagon decreases like
$\left(  \ln V_{R}\right)  /V_{R}$ with increasing $V_{R}$. In the equal mass
case this yields a value $\hat{V}_{R}\approx6.71197mc^{2}$, where
$m=M_{\text{tot}}/3$. For $m=M_{\text{tot}}/2$\ we obtain $\hat{V}_{R}%
\approx6.886682mc^{2}$ which is the maximal possible critical value of the
potential, and in the limits $m\longrightarrow0,M_{\text{tot}}$ we find
$\hat{V}_{R}\longrightarrow M_{\text{tot}}$. A plot of the critical values of
the potential as a function of $m_{j}$is given in fig.~\ref{fig:Vcritvsm}.

\begin{figure}[ptb]
\begin{center}
\epsfig{file=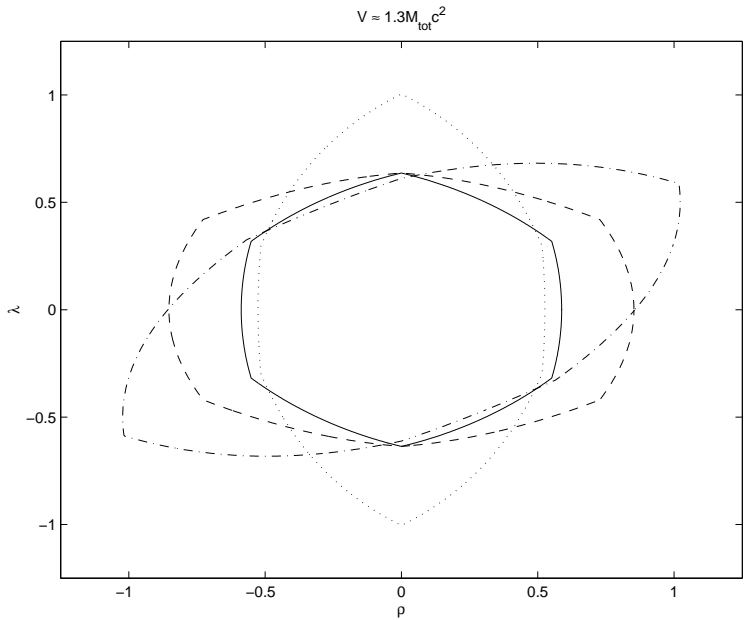, width=0.6\linewidth}
\end{center}
\caption{Cross-sections of the relativistic potential at $V\approx
1.3M_{tot}c^{2}$ for different ratios of the particle masses. The
correspondence between line style and ratio is the same as in
Figure~\ref{fig:newtpot}. The deformation of the potential due to changing the
mass ratio is the same as in the Newtonian case.  The $\rho$ and $\lambda$ are dimensionless variables as in Figure~\protect\ref{fig:eqpot}.}%
\label{fig:relpot}%
\end{figure}

The overall shape of the relativistic potential is that of a distorted,
hexagonal carafe. The distortion of the relativistic potential for different
ratios of the particle masses is analagous to the Newtonian potential and can
be seen in Figure~\ref{fig:relpot}. 

\begin{figure}[ptb]
\begin{center}%
\begin{tabular}
[c]{cc}%
\epsfig{file=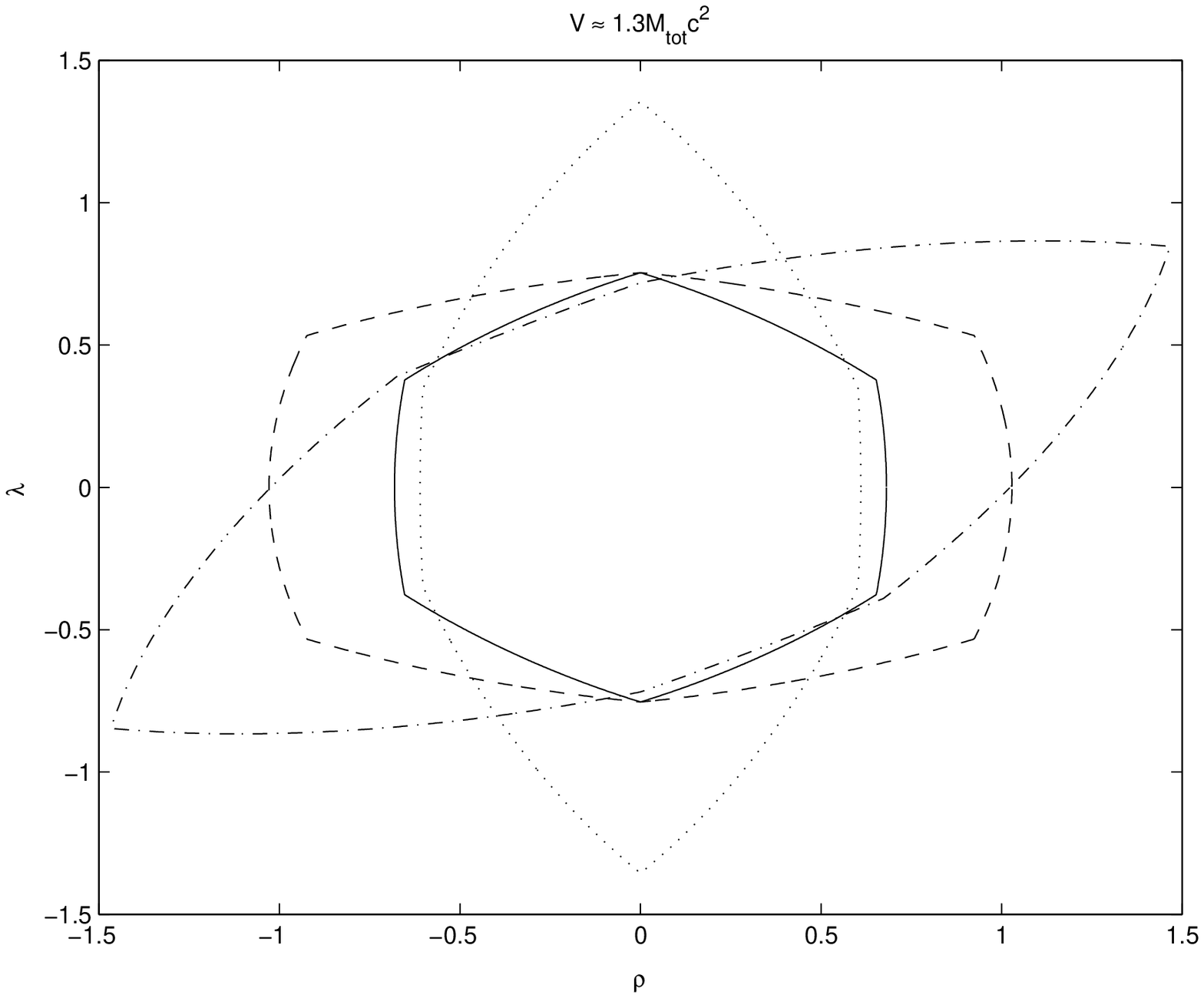, width=0.45\linewidth} & \epsfig{file=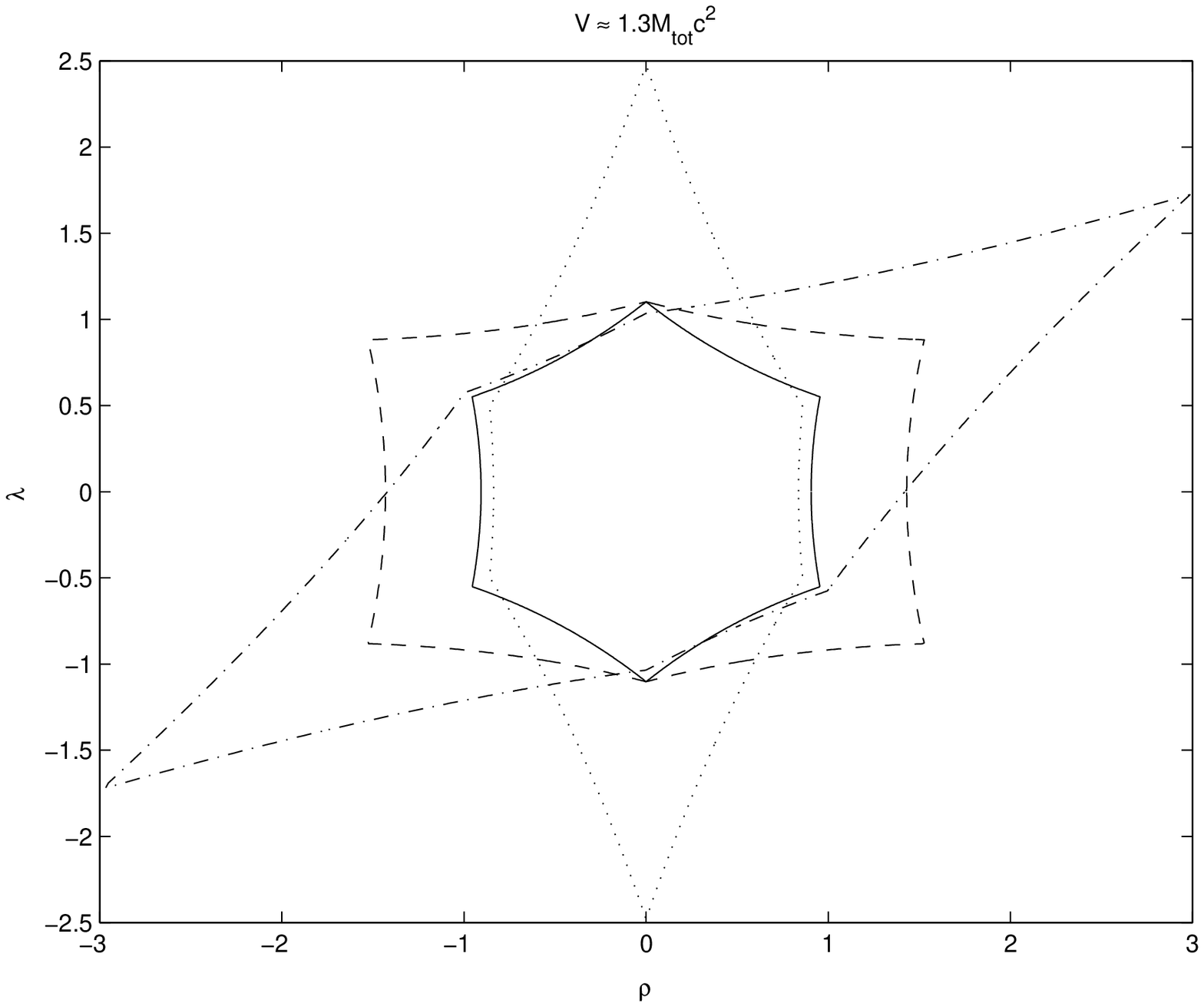,
width=0.45\linewidth}%
\end{tabular}
\end{center}
\caption{Cross-sections of both the untransformed (left) and transformed
post-Newtonian potentials at $V \approx1.3M_{tot} c^{2}$ for various mass
ratios. The correspondence between ratio and line is the same as in
Figure~\ref{fig:newtpot}.  The $\rho$ and $\lambda$ are dimensionless variables as in Figure~\protect\ref{fig:eqpot}.}%
\label{fig:pnpot}%
\end{figure}

The untransformed post-Newtonian potential shares similar features with the
relativistic potential in that the sides of the distorted hexagonal
cross-section are concave outward. However, as one might expect, the potential
increases less rapidly than the relativistic potential but still more rapidly
than the Newtonian potential at small values of $(\rho, \lambda) $.
Furthermore, the sides of the well continue to increase quadratically with
increasing $(\rho, \lambda)$ without the slope ever going to infinity as in
the relativistic case. Figure~\ref{fig:pnpot} shows a cross-section of the
untransformed post-Newtonian potential at a fixed value of $V$ for different
mass ratios. 

The transformed post-Newtonian potential has a much different character than
all of the potential energy functions discussed so far. The sides of the
distorted hexagon become convex and the vertices are always coincident with
the Newtonian potential at a fixed value of $V$. As $V$ increases, the sides
become more convex with respect to the Newtonian potential. Cross-sections of
this potential for different ratios of the particle masses at a fixed value of
$V$ can be seen in Figure~\ref{fig:pnpot}.

Finally, we note that the potential energy does not completely govern the
motion in the R and pN cases as it does in the N case due to the momentum
dependence of $V$ in the former cases. Consequently, such comparison of the
potentials is limited in the insight it can provide.

\section{Methods for Solving the Equations of Motion}

\label{sec:methsolv} The motion of the 3 particles under study is quite
complex so we have adopted several methods to study the equations of motion.
The most straight forward approach is to look at the position of each particle
with respect to the center of mass $\sum m_{a}z_{a}$ as a function of time
(where time will be explicitly defined shortly). Recall, however, that the
choice of the center of mass reference frame is arbitrary and is not
necessarily stationary to an observer as it is in the Newtonian case.

Under the change of coordinates (\ref{rholambdef}) and (\ref{prholambdef}),
the motion of the 3 particles is isomorphic to the motion of a single particle
moving in the hexagonal well potential (\ref{npot}) in the Newtonian case. The
situation is analogous in the fully relativistic and post-Newtonian cases
except that the potentials become momentum dependent. So, as in \cite{burnell}
and \cite{Bukta}, we focus on the trajectory of this particle, which we call
the `hex-particle', in the $\rho$-$\lambda$ plane as an alternate way to
analyze the motion.

As mentioned before, the bisectors (\ref{bisect1})--(\ref{bisect3}) correspond
to points where two of the particles are coincident. These bisectors divide
the $\rho$-$\lambda$ plane into 6 sextants corresponding to the 6 different
configurations the 3 particles can assume. So, when the hex-particle moves
from one sextant to the next, this corresponds to two particles passing
through each other. In the equal mass case, all 6 sextants are equivalent, and
in the unequal mass case, opposite sextants correspond to the opposite
configuration of particles (\textit{i.e.} $(1, 2, 3) \to(3, 2, 1)$). Further
symmetries exist when two particles have the same mass.

An analogous system was studied by Lehtihet and Miller \cite{LMiller,
MillRavi} who demonstrated that a self-gravitating Newtonian system of 3
particles in (1+1) dimensions \emph{with collisions} is equivalent to the
motion of a particle in a uniform gravitational field colliding elastically
with a wedge. The existence of particle collisions in our study of the N
system would correspond to the hex-particle being confined to a single sextant
where it would reduce to the particle-wedge system. In this particle-wedge
system, the equations of motion can be integrated between collisions of the
particle with the wedge and a discrete mapping describing the radial and
angular velocity of the particle at each collision can be used to describe the
motion. The simplification to a discrete mapping allows one to calculate fixed
points in phase space and evaluate their stability much more easily.
Unfortunately, the equations of motion for the pN and R systems are much more
complex than in the Newtonian case and it is not clear how to create a
discrete mapping between particle collisions.

Following \cite{LMiller} and \cite{burnell} we define two types of motion: $A$
motion, where two particles cross twice in succession; and $B$ motion, where a
single particle crosses each of the other two in succession. In $(\rho
,\lambda)$ space, after the hex-particle has just crossed one of the
bisectors, $A$ motion would correspond to a crossing of the same bisector
while $B$ motion would correspond to it crossing a different bisector. In this
way, one can describe the trajectory of the hex-particle as a succession of
$A$ and $B$ motions and develop a `symbol sequence' for a given trajectory. To
simplify the notation, we use exponents to denote a number of repeats of a
given type of motion so that the symbol sequence takes the form $\prod
_{i,j,k}\left(  A^{m_{i}}B^{n_{j}}\right)  ^{l_{k}}$ where $l_{k},m_{i}%
,n_{j}\in Z^{+}$ and $l_{k}$ is possibly infinite, in which case we denote it
by an overbar (\textit{i.e.} $\lim_{c\rightarrow\infty}(A^{a}B^{b})^{c}%
\equiv\overline{A^{a}B^{b}}$). Since the type of hex-particle motion at a
given bisector depends on the previous bisector, we avoid ambiguity by saying
that the first bisector crossing of the hex-particle is undefined, and the
symbol sequence begins at the second crossing. \ To aid in understanding this
nomenclature we have listed the symbol sequence in the captions of all
configuration space trajectories where the trajectory is easy to follow. \bigskip

The above methods allow us to study and classify individual trajectories of
the hex-particle in the 4 dimensional $(\rho,\lambda,p_{\rho},p_{\lambda})$
phase space. In order to study some of the global structure of this phase
space, we construct Poincar\'{e} maps. Since all of the Hamiltonians under
study are time independent, the total energy of the system is a constant of
motion and so the motion at a given energy is confined to a 3 dimensional
hypersurface in phase space. We can further reduce this to 2 dimensions by
plotting the radial momentum $p_{R}$ and the square of the angular momentum
$p_{\theta}^{2}$ of the hex-particle each time it crosses one of the
bisectors, as in \cite{LMiller} and \cite{burnell}. This is known as the
surface of section, or Poincar\'{e} map.

In the equal mass case, all bisectors are equivalent and so $p_{R}$ and
$p_{\theta}^{2}$ at each bisector can be plotted on the same surface of
section, as in \cite{burnell}. When the masses are unequal, this procedure is
not possible and one must distinguish between the different bisectors and the
directions in which they cross. We have chosen to plot points on the
Poincar\'e map each time the hex-particle crosses the $\rho= 0$ boundary in
the positive angular direction (\textit{i.e.} when $p_{\theta}> 0$). In the
case when $m_{1} = m_{2}$, particles 1 and 2 are indistinguishable and we may
also plot $(p_{R}, p_{\theta}^{2})$ each time the hex-particle crosses the
$\rho= 0$ bisector in the negative angular direction ($p_{\theta}< 0$). Due to
the nature of the Hamiltonian phase space, the different surfaces of section
corresponding to the different bisectors and directions contain the same
information and no generality is lost in making the above choice.

Since we were unable to find a closed form solution to either the relativistic
determining equation (\ref{Htrans}) or the equations of motion (\ref{zdot})
and (\ref{pdot}), it was necessary to employ numerical techniques to study the
motion. Using a Matlab integration routine (\texttt{ode15s}) we were able to
solve the equations of motion in the N, pN, and R systems.

The ode15s routine uses a variable order method for solving stiff differential
equations \cite{numerical}. In order to control computational errors, we
imposed absolute and relative error tolerances in the numerical routine of
$\epsilon_{abs}=\epsilon_{rel}=10^{-8}$ so that the estimated error in each of
the dynamical variables $\rho(i),\lambda(i),p_{\rho}(i)$, and $p_{\lambda}(i)$
at each step $i$ in the numerical integration is
\begin{equation}
\epsilon(i)\leq\max\left(  \epsilon_{rel}\left|  y(i)\right|  ,\epsilon
_{abs}\right)
\end{equation}
where $y(i)$ represents a generic, dimensionless dynamical variable at time
step $i$. These dimensionless variables will be introduced shortly. \bigskip

Furthermore, we periodically checked that the total energy of the system
remained constant to ensure that the solution was stable and physically correct.

We found (both in this study and in ref. \cite{burnell}) that the numerical
precision available to the computer did not allow the integration routine to
solve the equations at energies approximately $H\geq2M_{tot}c^{2}$. We were
unable to find a numerical integration routine that could integrate the
equations of motion in this energy regime so the dynamics of the system at
high energies still remains an open problem.

Furthermore, when we integrate the pN equations of motion, we find that the
resulting energy of the numerical solution does not remain constant in time,
despite the fact that the Hamiltonian~(\ref{pnham}) describes a conservative
system. The variation in energy becomes greater as the differences between the
masses increases. For the case when all masses are equal, this variation is on
the order of the imposed numerical error tolerances and can be ignored. A
description of the dynamics of the pN system in the equal mass case is given
in \cite{burnell}. The variation in energy increases drastically when we
change the ratio of the masses even by a small amount. For example, when we
integrate the equations of motion in the case where the mass of one particle
is half that of the other two, we see a variation in the total energy on the
order of $10^{-2}M_{tot}c^{2}$ over the duration of the trajectory. The cause
of this energy variation is unclear but its magnitude is clearly too large to
ignore. Due to this energy fluctuation, the numerical solutions to the
post-Newtonian equations of motion that we obtained in the unequal mass case
are clearly unphysical and will not be presented in this paper.

We cast the expressions for the Hamiltonian and the equations of motion in the
different systems in a dimensionless form using the coordinates $\hat{z_{i}}$
and $\hat{p_{i}}$, given by
\begin{align}
z_{i}  &  =\frac{4}{\kappa M_{tot}c^{2}}\hat{z_{i}}\label{zhat}\\
p_{i}  &  =M_{tot}c\hat{p_{i}}. \label{phat}%
\end{align}
We then express the dimensionless Hamiltonian as
\begin{equation}
\eta=\frac{H}{M_{tot}c^{2}}-1
\end{equation}
so that $\eta=0$ corresponds to $H$ being equal to the total rest mass of the
system. The total, dimensionless energy for all systems is $\eta+1$ (recall
that we are assuming the Newtonian Hamiltonian has been rescaled so that the
zero point is the total rest-energy of the system). In this way, a single
value of $\eta$ corresponds to the same energy in all 3 systems.

The equations of motion then become
\begin{align}
\frac{\partial\eta}{\partial\hat{p_{i}}}  &  =\frac{1}{c}\frac{\partial
H}{\partial p_{i}}=\frac{4}{\kappa M_{tot}c^{3}}\frac{d\hat{z}_{i}}{dt}%
=\frac{d\hat{z}_{i}}{d\hat{t}}\label{pdotscale}\\
\frac{\partial\eta}{\partial\hat{z_{i}}}  &  =\frac{4}{\kappa M_{tot}^{2}%
c^{4}}\frac{\partial H}{\partial z_{i}}=-\frac{4}{\kappa M_{tot}c^{3}}%
\frac{d\hat{p_{i}}}{dt}=-\frac{d\hat{p_{i}}}{d\hat{t}} \label{zdotscale}%
\end{align}
where we recognize $\hat{t}$ as the dimensionless time unit, given as
\begin{equation}
t=\frac{4}{\kappa M_{tot}c^{3}}\hat{t} . \label{t-hat}%
\end{equation}
We refer to $\hat{t} = 1$ as one time step. $\hat{\rho}$, $\hat{\lambda}$,
$\hat{p}_{\rho}$, and $\hat{p}_{\lambda}$, the dimensionless counterparts of
$\rho$, $\lambda$, $p_{\rho}$, and $p_{\lambda}$ respectively, are defined as
in (\ref{rholambdef}) and (\ref{prholambdef}) using the hatted variables of
(\ref{zhat}) and (\ref{phat}). In the subsequent analysis, dimensionless
variables will be assumed unless otherwise stated.

\section{Solution to the Equations of Motion}

\label{sec:solvequ} In this section we present the results of our numerical
analysis of the equations of motion. In \S~\ref{sec:eqmassoln} we summarize
the equal mass results of \cite{burnell} then go on to present how the
dynamics change in the unequal mass case in \S~\ref{sec:uneqmass} and
\S~\ref{sec:globstruct}.

\subsection{Equal Mass Solution}

\label{sec:eqmassoln} The study of the N, pN, and R systems when all three
particles share the same mass revealed a large variety of different types of
trajectories. The different types of motion can be classified into 3 broad
categories which we call annulus, pretzel, and chaotic. Note that our naming
scheme is not standard in the literature of dynamical systems. Our
nomenclature was chosen because of its direct physical interpretation in terms
of the three particles (despite the fact that the terms annulus and pretzel
derive from the shape of the trajectories in the $rho$-$lambda$ plane).

Annulus trajectories correspond to the hex-particle never crossing the same
bisector twice in a row resulting in an orbit about the origin of the $(\rho,
\lambda)$ plane. The symbol sequence for all annulus orbits is $\overline{B}$.
In terms of particles, these trajectories represent all motions in which no
two particles cross successively. Most of the trajectories in this class never
exactly repeat themselves after any number of orbits about the origin. The
result is a densely filled region of the $(\rho, \lambda)$ plane circling the
origin. All of these trajectories form closed loops on a Poincar\'e map.

Pretzel trajectories are so named because of the complex patterns they make
when plotted in $(\rho, \lambda)$ coordinates. Symbolically, these
trajectories fall into two distinct classes: 1) \textit{regular} trajectories,
which are denoted by some repeating pattern of $A$'s and $B$'s (\textit{e.g.}
$\overline{A^{2}B^{12}A^{5}B^{3}}$) and 2) \textit{quasi-regular}
trajectories, represented by some repeating pattern of $A$'s and $B$'s with
extra $A$ motions occasionally occurring on each repetition of the pattern
(\textit{e.g.} $A^{3}(A^{2}B^{6})^{3}A^{2}(A^{2}B^{6})^{11}$\ldots). As in the
annulus case, most pretzel trajectories never exactly repeat themselves and
densely fill a region of phase space. Pretzel trajectories appear either as a
series of small enclosed loops or as a series of disconnected lines on a
Poincar\'e map.

Chaotic trajectories are those that eventually cover all allowed regions of
phase space and are denoted symbolically by an apparently random sequence of
$A$'s and $B$'s. Since chaotic trajectories erratically cover a large region
of phase space, they appear as densely filled regions on a Poincar\'{e} map.
In all three systems there is a region of chaos separating the annulus and
pretzel regions on the surface of section \cite{burnell}.

A comparison between the relativistic and Newtonian systems reveals
differences in the trajectories as $\eta$ increases. In general, the particles
in the relativistic system cross each other at a higher frequency than in the
Newtonian case for the same value of $\eta$. The structure of the relativistic
Poincar\'{e} maps at all energies attainable were similar to the Newtonian
ones except for a shifting of all trajectories to one side. Remarkably, this
structure remained stable up to the values of $\eta$ that were attainable
despite the high degree of non-linearity in the equations of motion.

Furthermore, for all trajectories studied in all 3 systems, $B$ motion always
occurred in multiples of 3 \cite{burnell}. That is, all symbol sequences were
of the form $\prod_{i,j,k}(A^{m_{i}}B^{3n_{j}})^{l_{k}}$ so that any time a
single particle, say particle 1, crossed the other two in succession,
particles 2 and 3 always crossed next before meeting particle 1 again.

\subsection{Unequal Mass Trajectories}

\label{sec:uneqmass} In order to study the effects of changing the relative
masses of the particles, we adopt the parameter $\alpha$ whenever two masses
are equal so that $m_{1} = m_{2} = \alpha m_{3}$. When all three masses are
unequal, we will describe the relative masses as a ratio (\textit{i.e.}
$m_{1}:m_{2}:m_{3} = 1:2:3$).

\begin{figure}[ptb]
\begin{center}
\epsfig{file=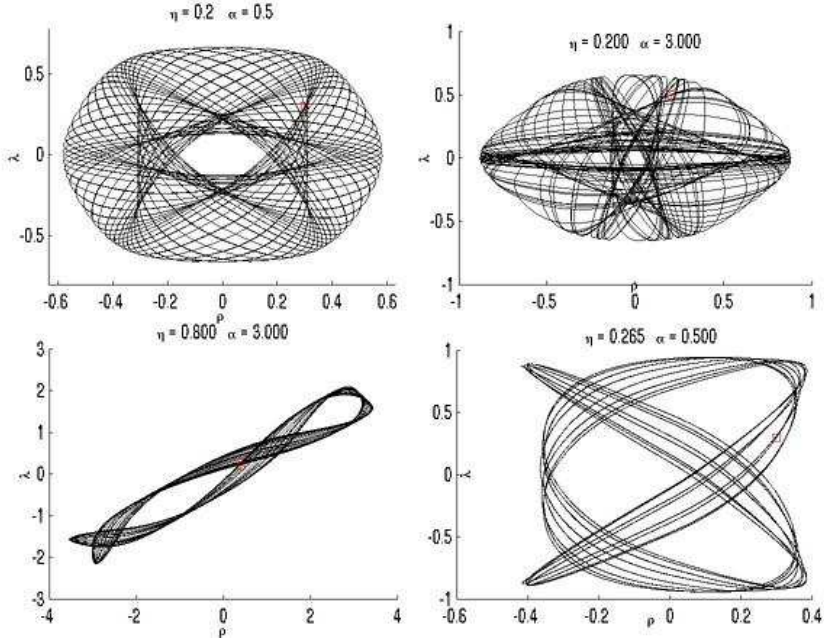, width=0.7\linewidth}
\end{center}
\caption{Examples of trajectories in the N system for different values of
$\eta$ and $\alpha$. Each trajectory was run for 150 time steps. The small box
indicates the starting position of the trajectory. Proceeding clockwise from
the top left plot the symbol sequences are $\overline{B}$, $\overline{B}$,
$\overline{AB^{3}A^{2}B^{3}}$, and $B^{6}A\overline{B^{9}A}$.}%
\label{fig:exN}%
\end{figure}

\begin{figure}[ptbptb]
\begin{center}
\epsfig{file=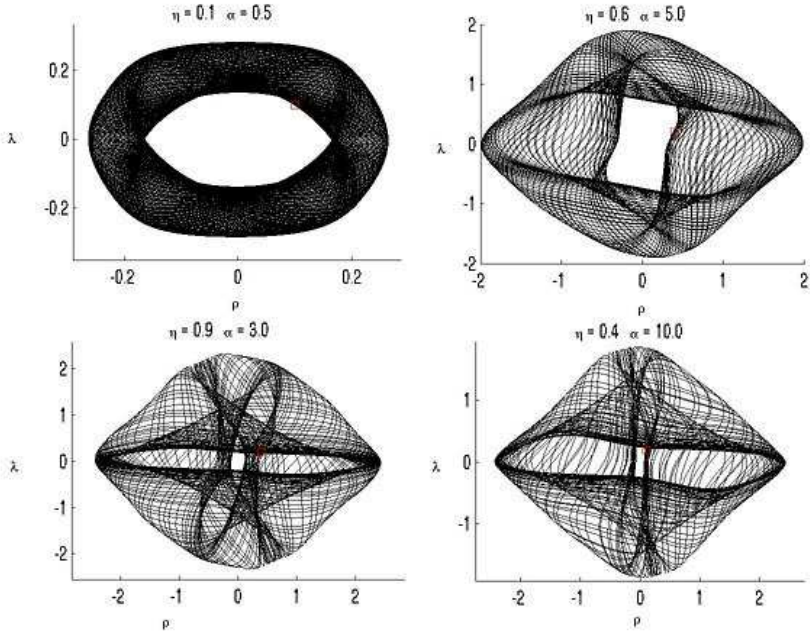, width=0.7\linewidth}
\end{center}
\caption{Examples of relativistic annulus trajectories for different values of
$\eta$ and $\alpha$. Notice the characteristic boxy shape at higher values of
$\alpha$. Each trajectory was run for 200 time steps. All trajectories have
the symbol sequence $\overline{B}$.}%
\label{fig:exannR}%
\end{figure}

\begin{figure}[ptbptbptb]
\begin{center}
\epsfig{file=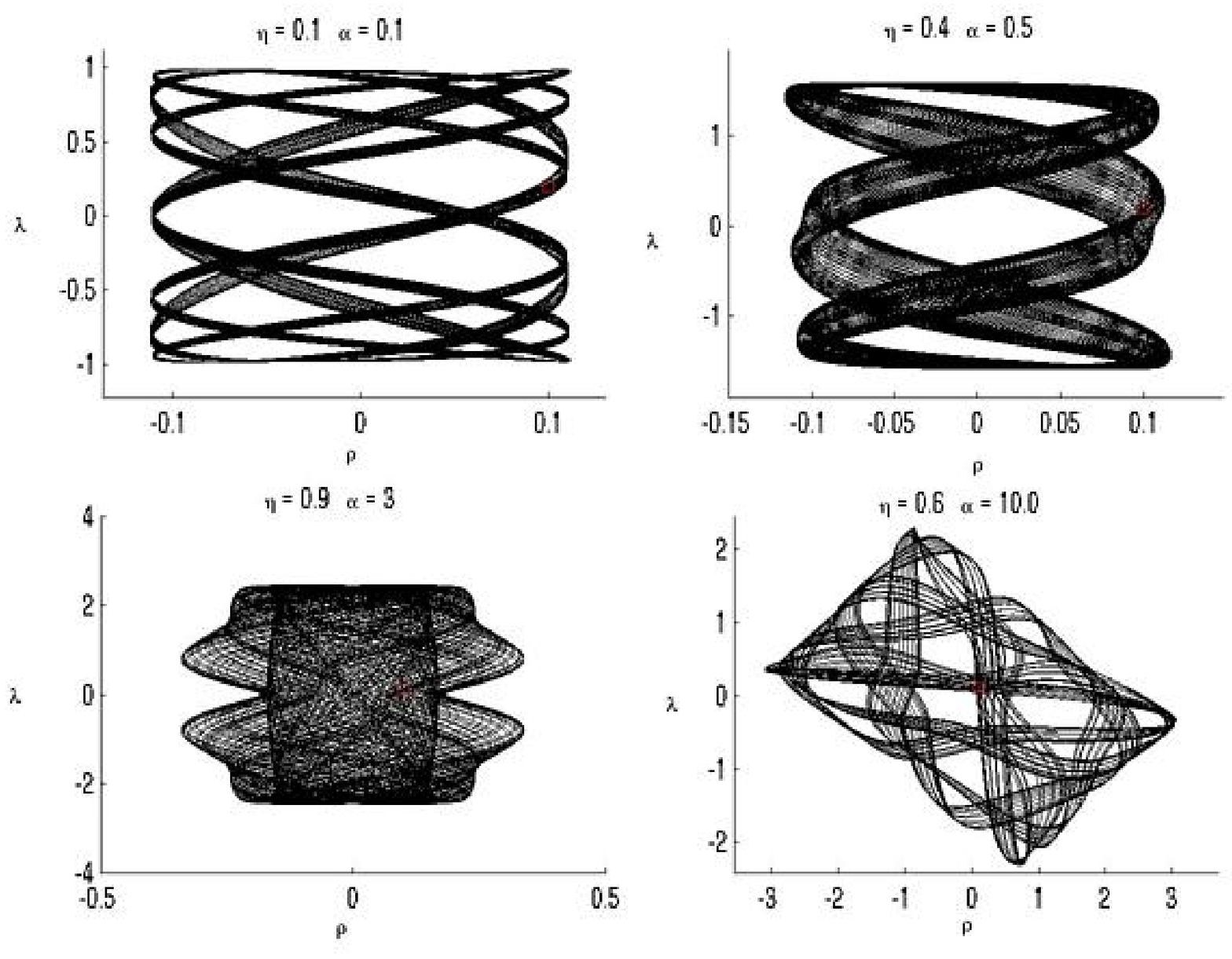, width=0.7\linewidth}
\end{center}
\caption{Examples of relativistic pretzel trajectories for different values of
$\eta$ and $\alpha$. Each trajectory was run for 200 time steps. The top right
plot has symbol sequence $\overline{A^{2}B^{3}}$.}%
\label{fig:expretR}%
\end{figure}

In the case when two masses are equal ($\alpha\neq1$), we find the same
diversity of trajectories in the $(\rho,\lambda)$ plane as in the $\alpha=1$
case \cite{burnell}. Figure~\ref{fig:exN} shows this diversity for different
values of $\alpha$ in the N system while Figures~\ref{fig:exannR}
and~\ref{fig:expretR} show examples of annulus and pretzel trajectories
respectively for the R system. Similar trajectories are obtained when all
three masses are unequal. 

In terms of the hex-particle moving in the $\rho$-$\lambda$ plane, we did not
find a significant difference between the equal and unequal mass trajectories
besides a general distortion of the annulus orbits as the difference in masses
increases. For instance, in the equal mass case \cite{burnell}, all of the
annulus orbits were generally hexagonal about the origin. When the mass of one
particle is larger than the rest, these annuli take on a more box-like shape,
as can be seen in Figure~\ref{fig:exannR}. Besides a general distortion, we
did not find any novel types of motion that were not seen in the equal mass
case. Since the qualitative aspects of the motion in the $\rho$-$\lambda$
plane does not reveal much about the underlying physics, we will forgo any
further discussion on this matter.

Another effect of changing the difference between the particle masses is that
the ratio between the number of annulus trajectories compared to the number of
pretzel orbits at a given energy decreases. That is, as $\alpha$ decreases, we
find fewer and fewer initial conditions that give annuli compared to initial
conditions that produce pretzels. The reason for this is best demonstrated by
looking at the motion of the particles as a function of time.

\begin{figure}[ptb]
\begin{center}
\epsfig{file=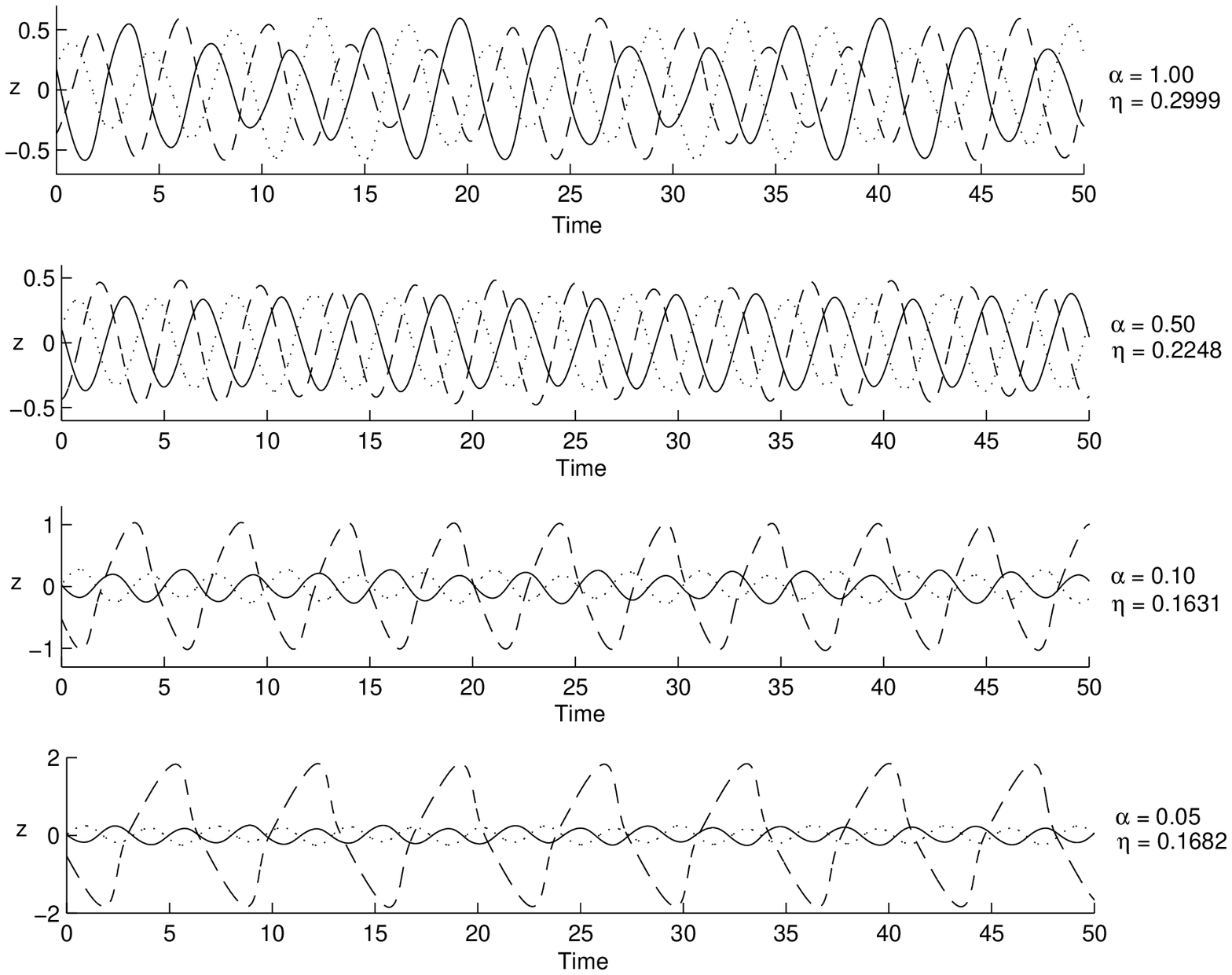, width=1\linewidth}
\end{center}
\caption{The relative position of each particle with respect to the center of
mass is plotted as a function of time for various values of $\alpha$ in the R
system. The particles 1, 2, and 3 have relative masses in the ratio
$1:1:\alpha$. Solid line - particle 1, dotted line - particle 2, and dashed
line - particle 3. Each plot uses the same initial values of $(\rho
,\lambda,p_{}\rho,p_{}\lambda)$ but the total energy $\eta+1$ is fixed by the
energy constraint (\ref{Htrans}). The top two plots display annulus motion
($\overline{B}$) while the bottom two are classified as pretzel trajectories
($\overline{\left(  B^{6}A\right)  ^{7}B^{3}}$ and $\overline{A^{2}%
B^{3}\left(  AB^{3}\right)  ^{5}}$ respectively).}%
\label{fig:tzlowalphaR}%
\end{figure}Figure~\ref{fig:tzlowalphaR} plots the relative motion of the
particles for decreasing values of $\alpha$ in the R system for a specific set
of initial conditions. We see that, at equal mass ($\alpha=1$), a single
particle alternately crosses the other two without ever crossing the same
particle twice, indicative of annulus motion. However, as $\alpha$ decreases,
the mass of particle 3 decreases and so its frequency of oscillation decreases
while its amplitude increases with respect to the other two.

\begin{figure}[ptb]
\begin{center}
\epsfig{file=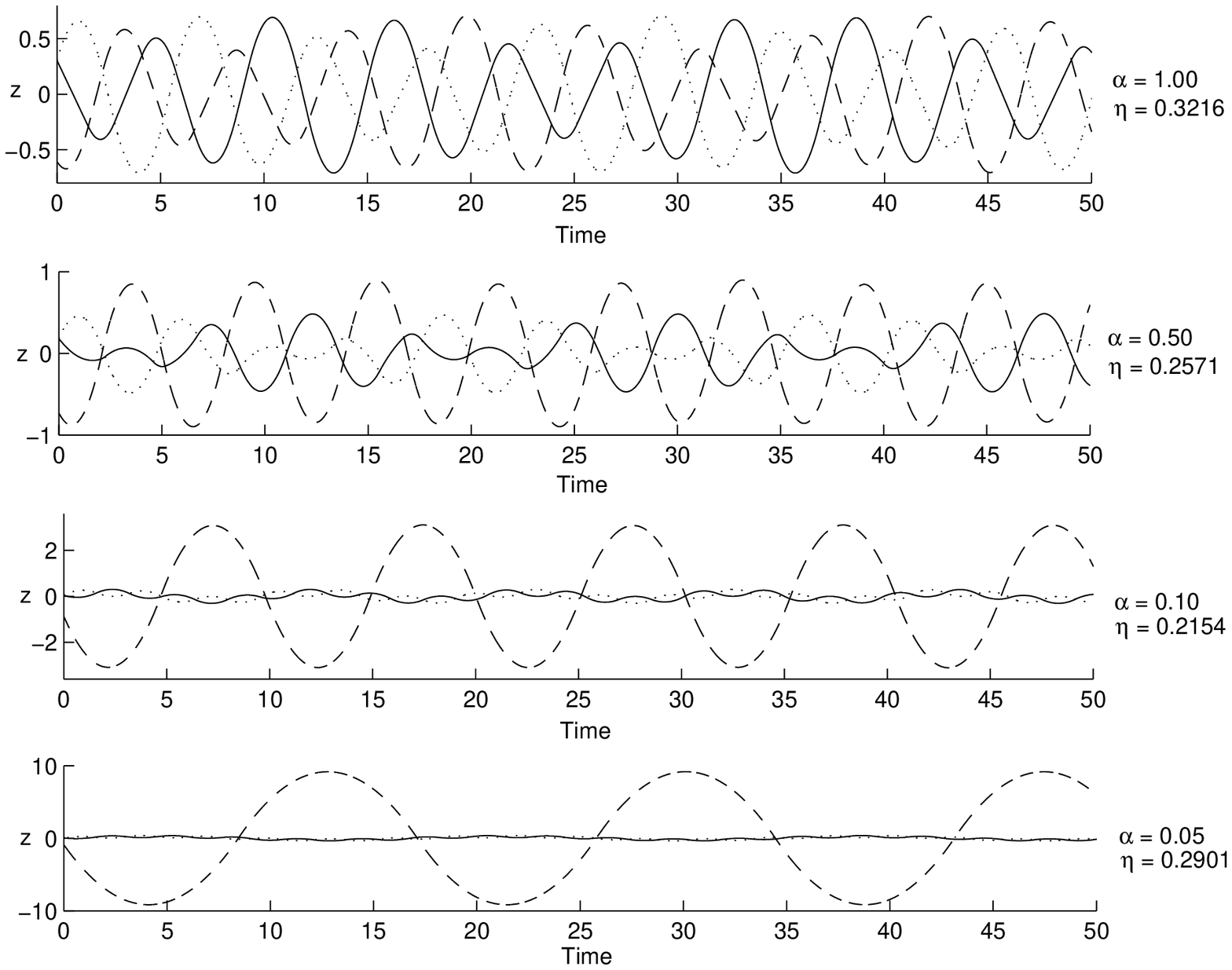, width=1\linewidth}
\end{center}
\caption{The relative positions of each particle with respect to the center of
mass as a function of time in the N system. These plots were created using the
same procedure as in Figure~\ref{fig:tzlowalphaR} and follow the same
conventions except that equation (\ref{Hnewt}) is used to fix the value of the
total energy $\eta$. The first is an annulus trajectory ($\overline{B}$) while
the remaining are pretzels ($\overline{B^{9}A}$, $\overline{\left(  B^{3}%
A^{2}\right)  ^{4}B^{3}A^{3}}$, and $\overline{\left(  B^{3}A^{5}\right)
^{2}B^{3}A^{4}}$ from top to bottom).}%
\label{fig:tzlowalphaN}%
\end{figure}

In effect, we see that the two massive particles gravitationally bind together
more tightly as the difference between their mass and the mass of the third
particle increases. Eventually, this binding becomes so tight that the two
massive particles are forced to execute an additional $A$ motion before
crossing the third particle and, hence, there is a transition from annulus
type motion to pretzel type motion. This behavior, while expected for the
Newtonian system (Figure~\ref{fig:tzlowalphaN}), is also present in the
relativistic case. \bigskip

As the mass difference increases, it is much more difficult to set up initial
conditions at a given energy such that particle 1 and 2 do not cross more than
once during one of particle 3's long period oscillations. This effect is also
seen in the Newtonian system, as shown in Figure~\ref{fig:tzlowalphaN}. This
difference in the ratio of the number of annulus trajectories compared to
pretzel trajectories will be made more clear in the next section when we look
at the Poincar\'{e} maps. 

\begin{figure}[ptb]
\begin{center}
\epsfig{file=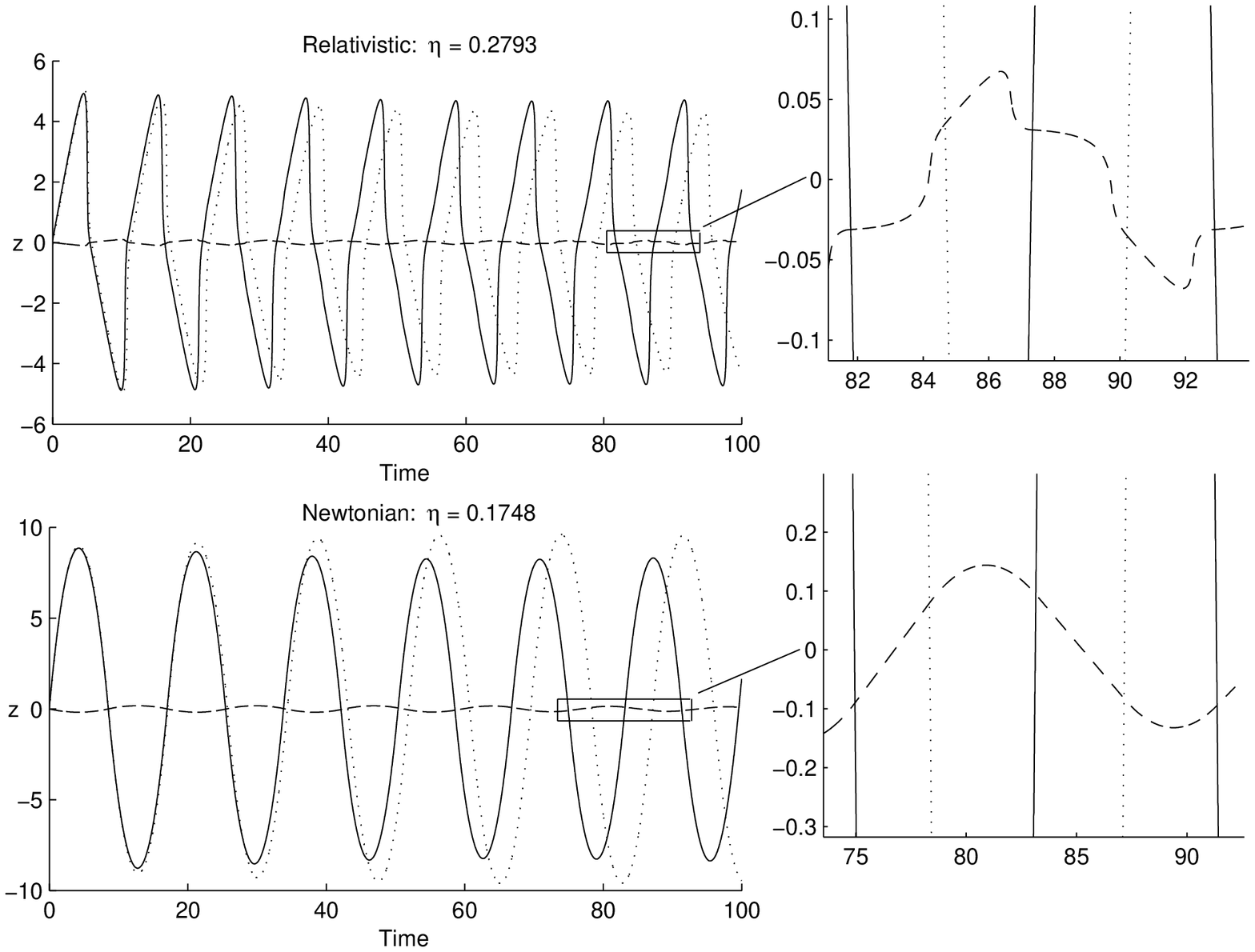, width=0.7\linewidth}
\end{center}
\caption{Relative motion of the particles with respect to the center of mass
plotted as a function of time for the R system (top) and the N system
(bottom). Both plots have mass ratios 1:1:100, or $\alpha= 100$. The lines are
as defined in Figure~\ref{fig:tzlowalphaR}. The insets show the small
perturbation in the motion of the large mass due to the crossing of the
smaller masses.}%
\label{fig:bigmass}%
\end{figure}

Using these position-time plots, it is interesting to explore the limit where
one mass is much greater and much smaller than the other two ($\alpha\gg1$ and
$\alpha\ll1$ respectively). The former case is shown in
Figure~\ref{fig:bigmass} for the N and R systems where $\alpha= 100$.
In both cases, as one would expect, we see the large mass barely
moves while the other two particles oscillate about it. The inset shows the
small perturbations to the motion of the larger mass caused by the passing of
the two particles. In the Newtonian case, the perturbation is very smooth and
regular while the perturbation in the relativistic case is more jerky and
erratic. That is, the velocity of the large mass in the R system increases
much more suddenly than in the N system when it encounters another, smaller mass.

\begin{figure}[ptb]
\begin{center}
\epsfig{file=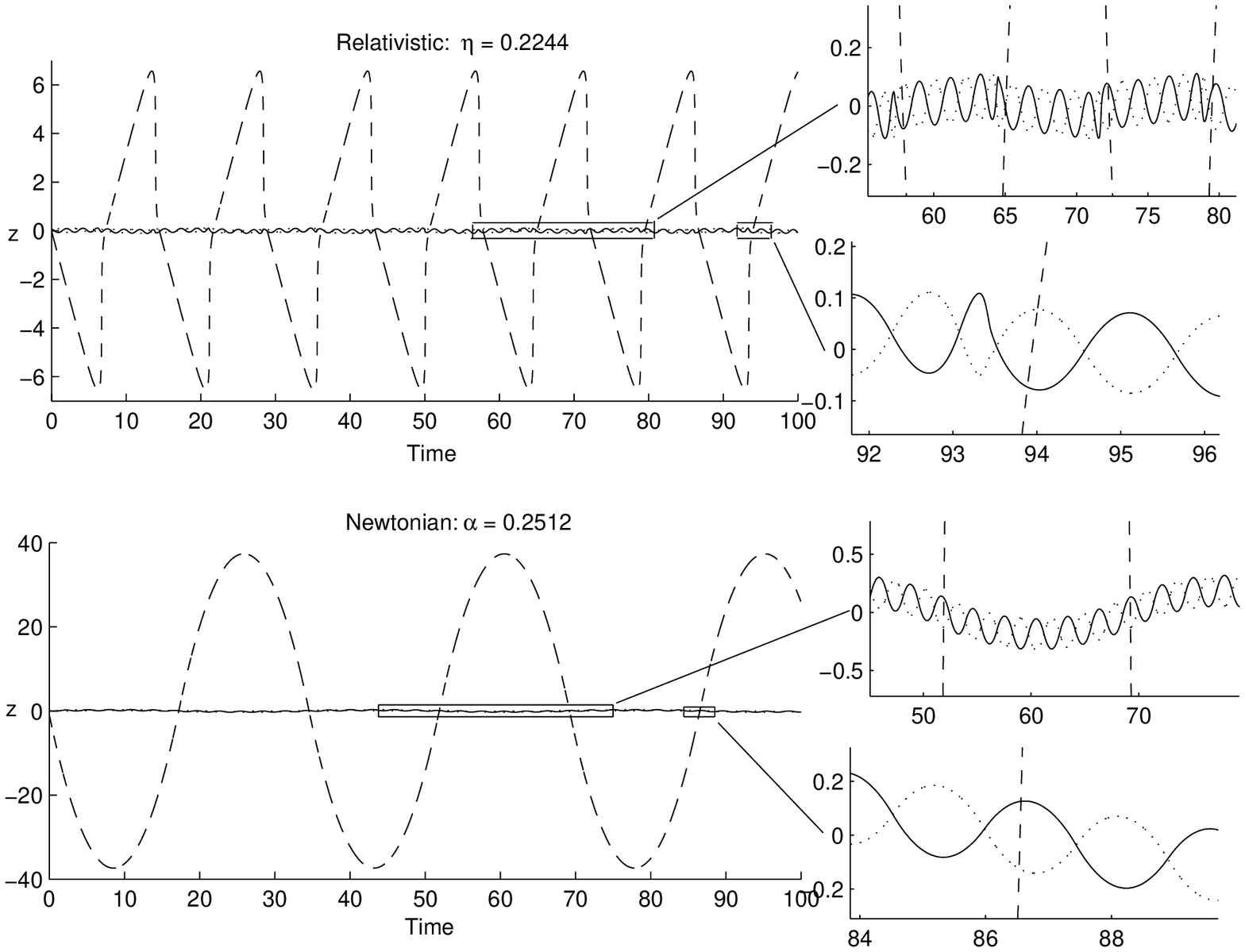, width=0.7\linewidth}
\end{center}
\caption{Relative motion of the particles for the case where the mass ratio is
1:1:0.01, or $\alpha=0.01$ for both the R (top) and N (bottom) systems. The
insets show the motion of the stable, two body sub-system made up of the two
heavy particles, as well as the effect of encounters with the light particle.}%
\label{fig:smallmass}%
\end{figure}Figure~\ref{fig:smallmass} shows the corresponding plots when
$\alpha=0.01$. We see that, in both cases, the two heavy particles form a
stable, 2-body sub-system while the third particle oscillates about their
center of mass. As seen in the upper most insets in both the R and N case, the
presence of the light particle has a weak gravitational effect, causing the
oscillatory motion of the center of mass of the two more massive particles.
Unlike the effect seen in Figure~\ref{fig:bigmass}, the perturbation of the
motion of the heavy particles due to the crossing of the light particle is
very small in the R system and almost imperceptible in the N system. The
reason for this is that the two heavy particles in the $\alpha=0.01$ case are
twice as massive as the single particle in the $\alpha=100$ and so the motion
of the 2-body subsystem is much more stable and less susceptible to harassment
from the weaker mass. However, the qualitative nature of the perturbation
remains the same as in the $\alpha=100$ case.

We also find that the amplitude of oscillations in the Newtonian system is
generally larger than in the relativistic system at corresponding values of
the total energy and that the frequency of oscillations is greater in the
relativistic case. These observations agree with the results found in the
equal mass case \cite{burnell}.

Finally, we note that, as in the equal mass case, we find that B motion always
comes in multiples of three. That is, the symbol sequence always takes the
form
\begin{equation}
\prod_{i,j,k}(A^{m_{i}},B^{3n_{j}})^{l_{k}} \label{symbseq}%
\end{equation}
for all values of $\alpha$ and $\eta$ that were studied. This extends our
hypothesis proposed in \cite{burnell}, that all trajectories in the R and N
systems, when translated into a symbol sequence, have the form (\ref{symbseq}%
), to also hold for all mass ratios of the 3 particles.

\subsection{Global Structure of Phase Space}

\label{sec:globstruct} By studying the 2 dimensional representations of phase
space represented in the Poincar\'e maps we were able to discover some
interesting global properties of both the N and R systems. We begin this
section by describing some of the basic features of the Poincar\'e plots and
then go on to discuss how the structure of phase space changes when the mass
ratio of the particles is changed. Our results will then be compared with
similar studies conducted previously.

\begin{figure}[ptb]
\begin{center}
\epsfig{file=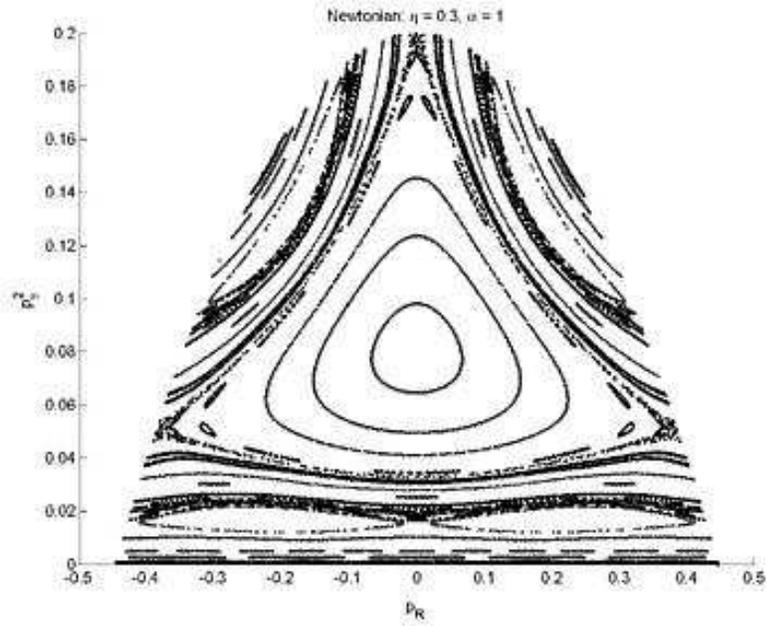, width=0.7\linewidth}
\end{center}
\caption{A Poincar\'e map of the Newtonian system when all of the particle
masses are equal.}%
\label{fig:poincNeq}%
\end{figure}

An example of a Poincar\'e map for the N system when all masses are equal is
shown in Figure~\ref{fig:poincNeq}. All points on this surface of section fall within a parabolic
region which is defined by the system's energy constraint. It was found in
\cite{burnell}, as mentioned previously, that the three types of motion,
annulus, pretzel, and chaotic fall into 3 regions on the surface of section.
Quasi-periodic annulus orbits form single closed loops about a stable fixed
point at which the motion is completely periodic. In Figure~\ref{fig:poincNeq}
this region of quasi-periodic annuli is located at the center of the plot
enclosed within the densely filled triangular shaped region. This densely
filled region is created by chaotic trajectories and separates the annulus
region from the pretzel region. All pretzel trajectories fall outside of this
chaotic region and form either a series of disconnected loops or a series of
disconnected lines.

\begin{figure}[ptb]
\begin{center}
\epsfig{file=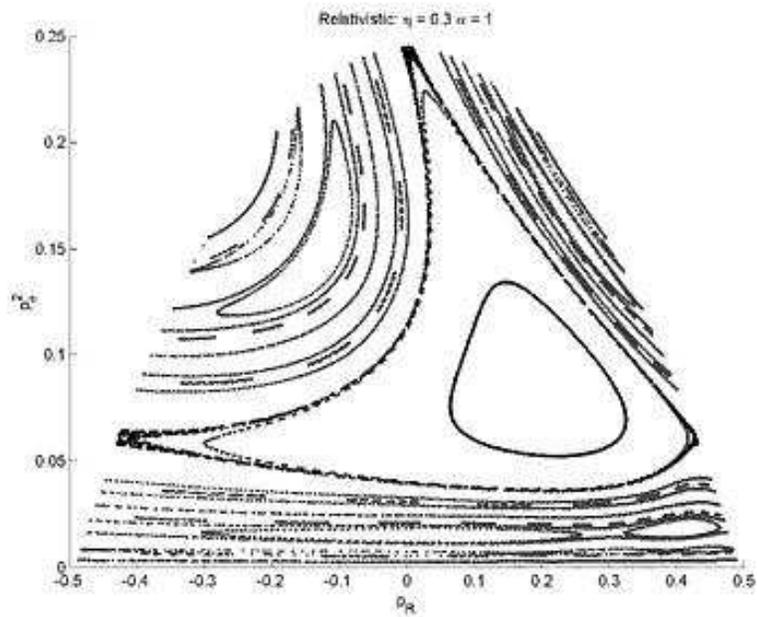, width=0.7\linewidth}
\end{center}
\caption{A Poincar\'e map of the relativistic system when the masses of all of
the particles are equal.}%
\label{fig:poincReq}%
\end{figure}

A similar segregation of the surface of section is also seen in the R system,
an example of which is shown in Figure~\ref{fig:poincReq}. The relativistic Poincar\'e map is strikingly similar to the
Newtonian one in Figure~\ref{fig:poincNeq}. The annulus region is shown as the
series of closed loops in the lower right portion, surrounded by the warped
chaotic region, which is further surrounded by the region of pretzel trajectories.

In general, the relativistic phase space is a warping of the corresponding
Newtonian space. As described in \cite{burnell}, this is due to the weaker
symmetry of the relativistic Hamiltonian compared to the Newtonian. As seen in
eq.~(\ref{Hnewt}), the N system is invariant under the symmetry $p_{i}%
\rightarrow-p_{i}$ and this is manifest in the symmetry about the $p_{R}=0$
axis in Figure~\ref{fig:poincNeq}. \bigskip The relativistic Hamiltonian,
determined by (\ref{Htrans}), is invariant under the symmetry $(p_{i}%
,\epsilon)\rightarrow(-p_{i},-\epsilon)$. Contrary to the Newtonian case, this
relativistic symmetry is not manifest in our surface of section. \bigskip

We find that the annulus and pretzel trajectories continue to fall into
similar regions, as described above, for all different mass ratios studied,
and that these two regions are always separated by a region of chaos. By
changing the mass ratio at a given value of the total energy, the size and
shape of the different regions change.

\begin{figure}[ptb]
\begin{center}%
\begin{tabular}
[c]{cc}%
\epsfig{file=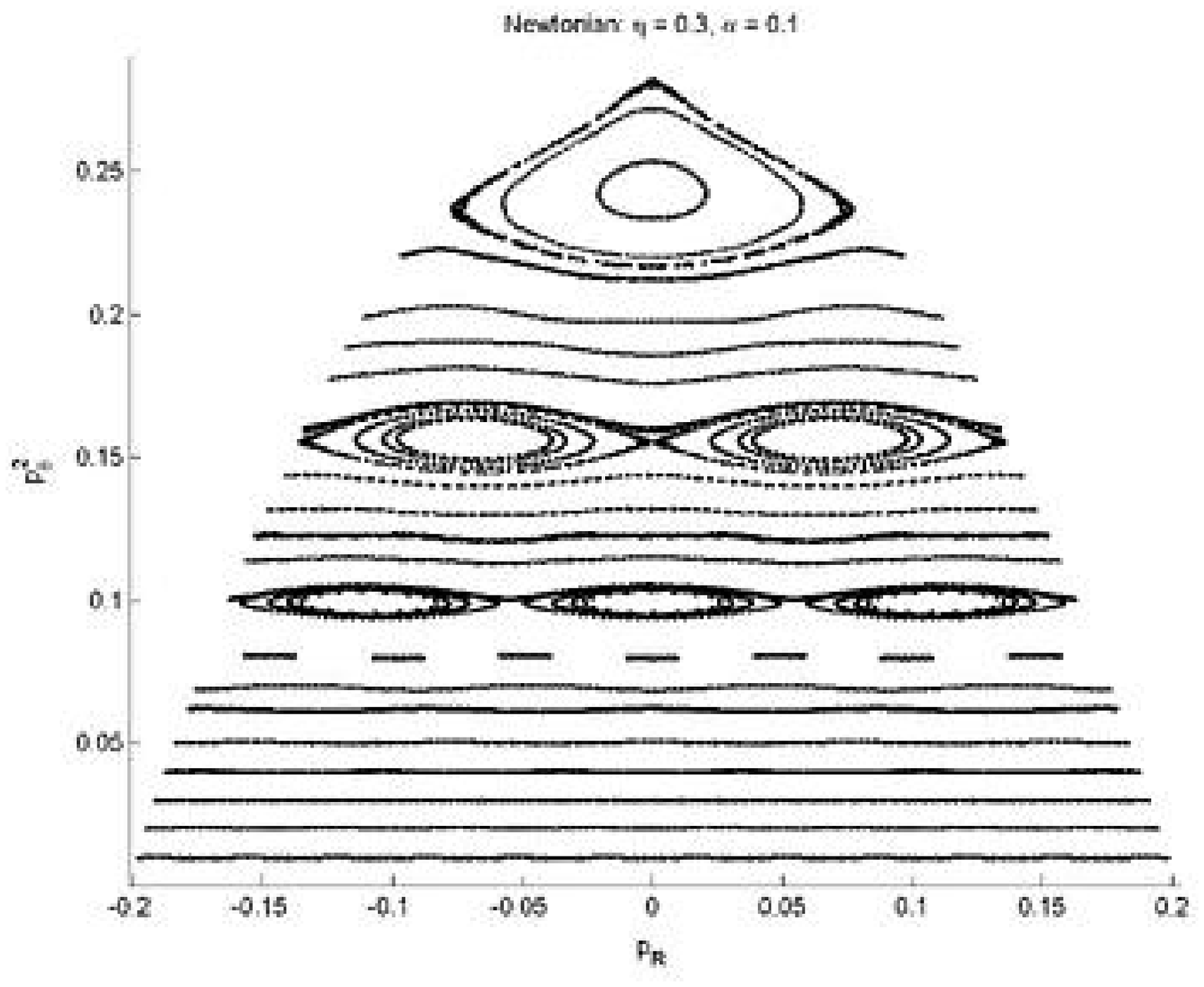, width=0.45\linewidth} & \epsfig
{file=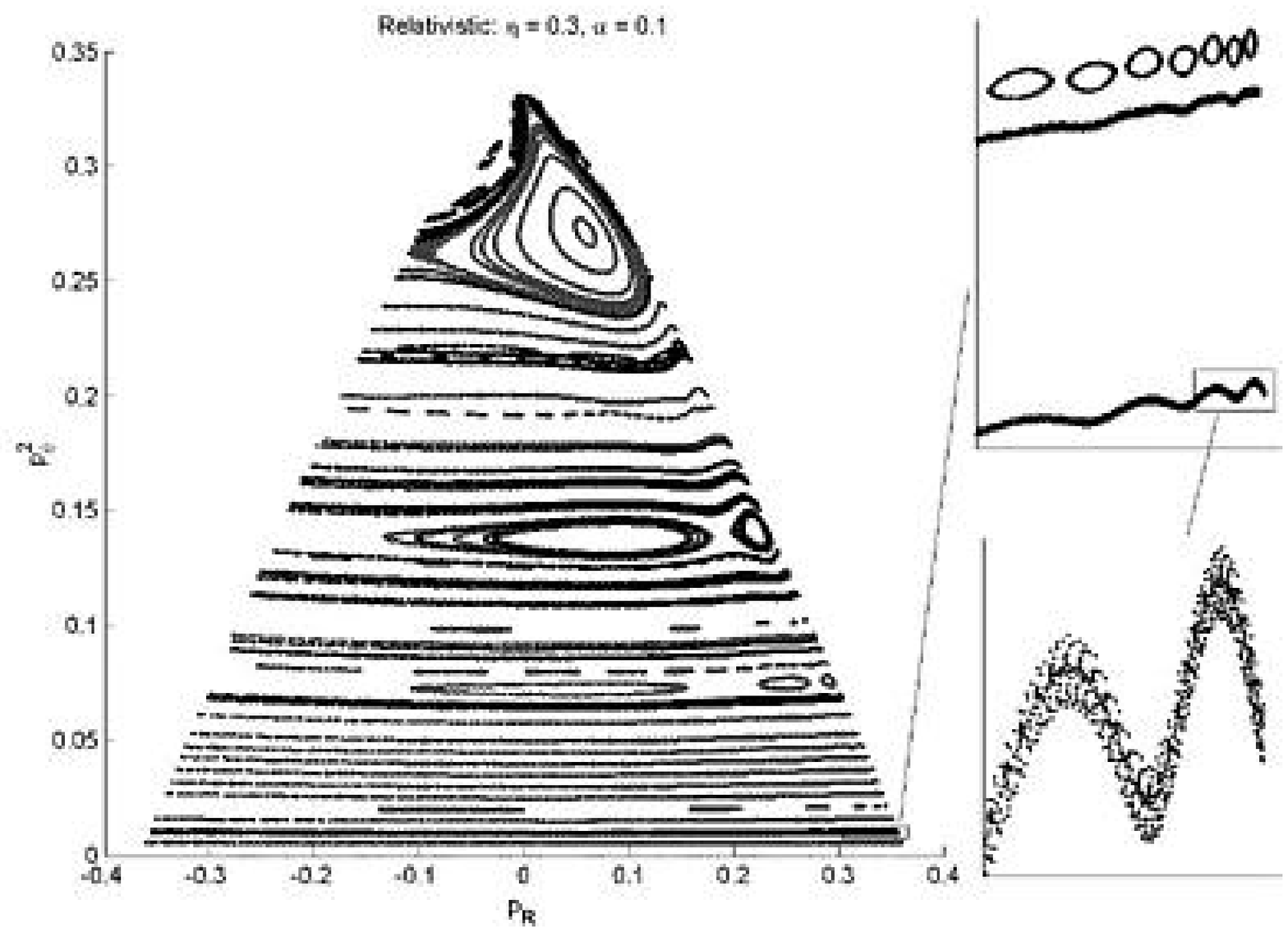, width=0.45\linewidth}%
\end{tabular}
\end{center}
\caption{Poincar\'e plots with $\alpha= 0.1$ for the Newtonian (left) and
relativistic (right) systems. The insets on the right show the onset of chaos
in the pretzel region.}%
\label{fig:poinclowalphaRN}%
\end{figure}

More specifically, by looking at the case where particles 1 and 2 share the
same mass, the annulus region becomes smaller and moves towards the top of the
allowed region of the surface of section with decreasing $\alpha< 1$ as can be
seen in Figure~\ref{fig:poinclowalphaRN} for $\alpha= 0.1$ in both the N and R
systems. This shrinking of the annulus region is a manifestation of the
effect discussed in \S\ref{sec:uneqmass} where annulus motion becomes more
difficult to attain when one particle is significantly less massive than the
other two.

\begin{figure}[ptb]
\begin{center}%
\begin{tabular}
[c]{cc}%
\epsfig{file=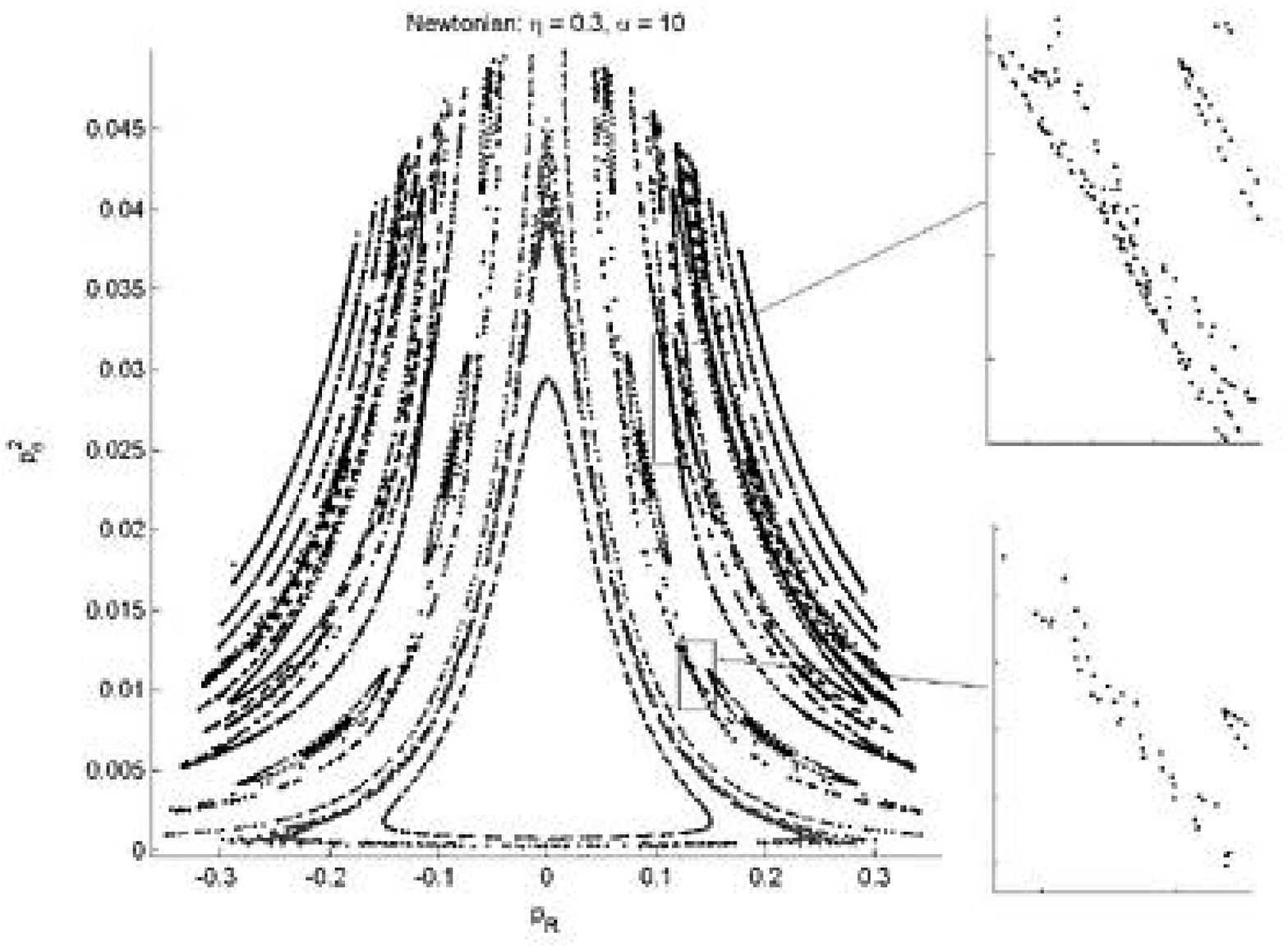, width=0.45\linewidth} & \epsfig
{file=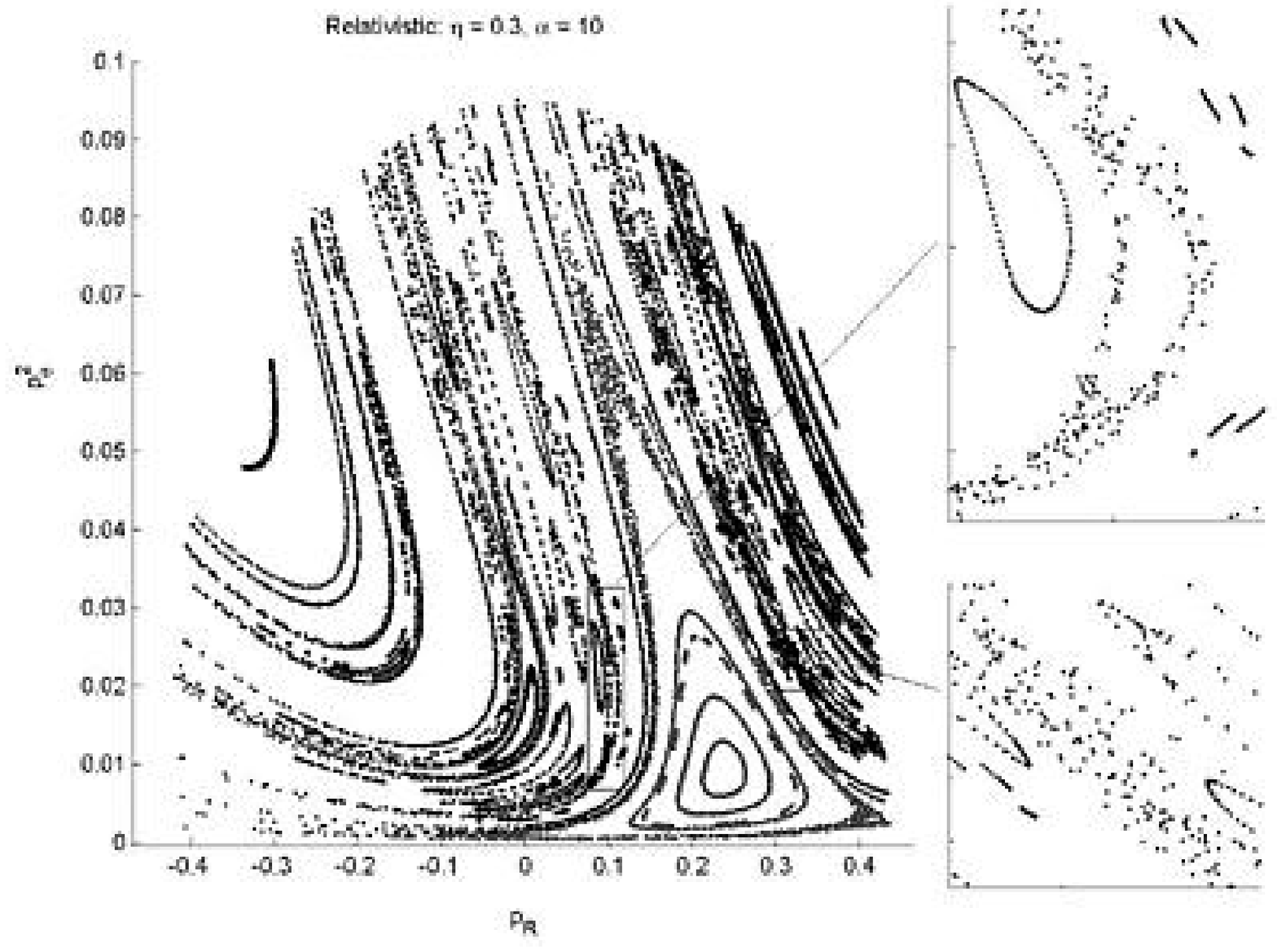, width=0.45\linewidth}%
\end{tabular}
\end{center}
\caption{Poincar\'{e} plots with $\alpha=10$ for the Newtonian (left) and
relativistic (right) systems. The insets show additional regions of chaos in
the pretzel region that are not present in the corresponding region on the
equal mass Poincare section.}%
\label{fig:poinchighalphaRN}%
\end{figure}

As $\alpha$ increases, the annulus region extends towards the lower region of
the plot, as shown in Figure~\ref{fig:poinchighalphaRN} for $\alpha=10$ in
both the R and N systems. Essentially, this means that one requires a lower magnitude of
angular momentum of the hex-particle to attain an annulus orbit in the $\rho
$-$\lambda$ plane. Since the gravitational attraction between the two light
particles is not very strong compared to their interaction with the heavy
particle, the light particles do not tend to oscillate about each other very
much but instead act like 2 separate 2 body systems with the heavy particle
taking the role of the second body, like a two-planet, one-dimensional solar
system. This situation is shown in Figure~\ref{fig:bigmass} for $\alpha=100$
and explains why there is no decrease in size of the annulus region with
increasing $\alpha$.

The symmetry about $p_{R} = 0$ present in Figures~\ref{fig:poinclowalphaRN}
and~\ref{fig:poinchighalphaRN} is really just an artifact of our choice of
surface of section. Recall that we chose to construct our Poincar\'e maps by
plotting a point each time the hex-particle crossed the $\rho= 0$ bisector,
or, equivalently, each time particles 1 and 2 crossed. The above figures were
constructed with $m_{1} = m_{2}$ and so the symmetries of the equal mass
system persist. If we were to have chosen a different bisector, all of the
features discussed above would remain (\textit{e.g.} shrinking, expanding of
annulus region) but these would not occur in the same sense and the plots
would not be as symmetrical.

\begin{figure}[ptb]
\begin{center}%
\begin{tabular}
[c]{cc}%
\epsfig{file=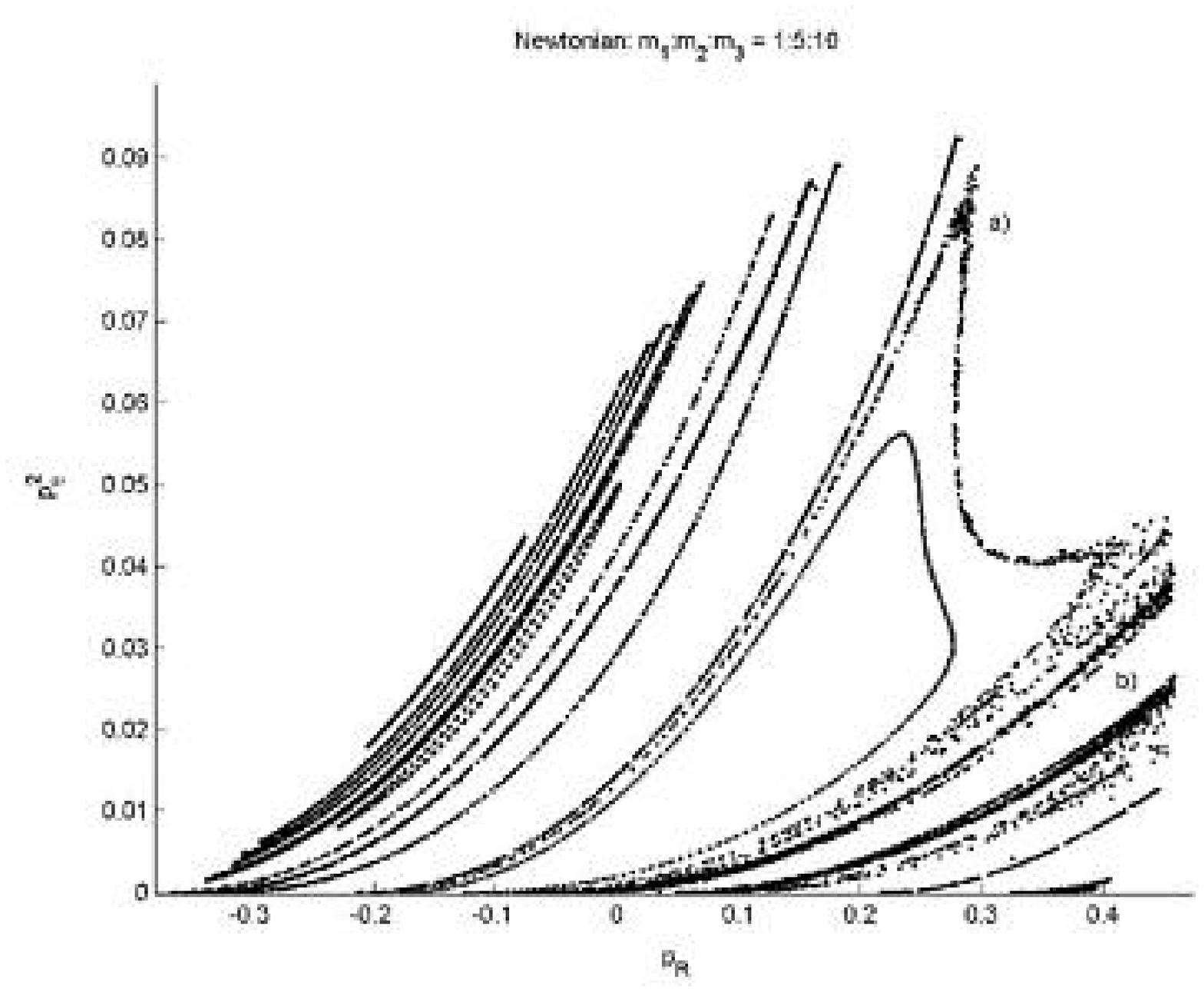, width=0.45\linewidth} & \epsfig
{file=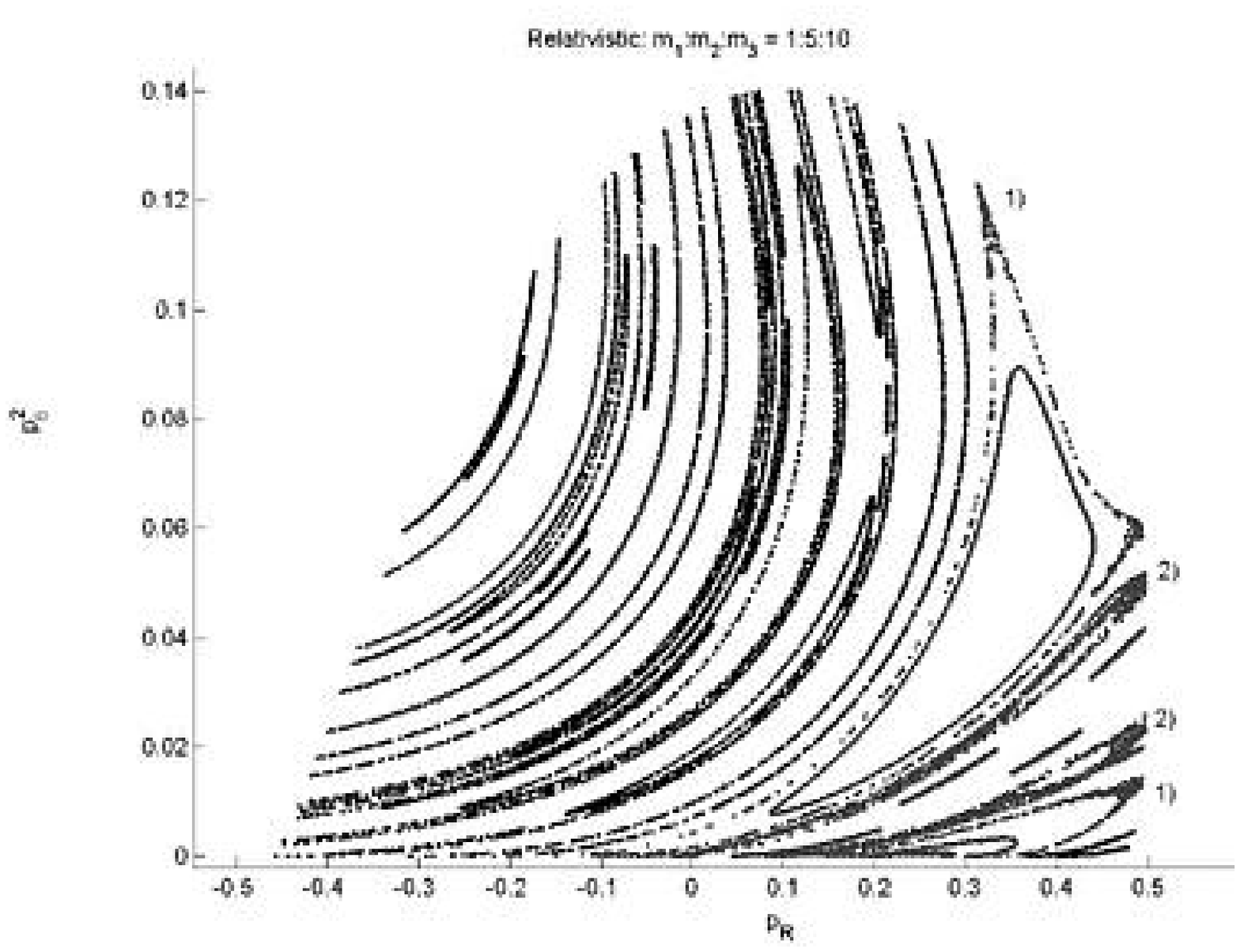, width=0.45\linewidth}%
\end{tabular}
\end{center}
\caption{Poincar\'{e} plots with a mass ratio of 1:5:10 for the Newtonian
(left) and relativistic (right) systems. On the left, a) marks the region of
chaos separating annulus trajectories (inside) and predominantly pretzel
trajectories (outside) while the densely filled area directly above and below
b) marks a new region of chaos amongst the pretzel trajectories. On the right,
the densely filled regions marked by a 1) were created by a single trajectory
separating the annulus and pretzel orbits while the chaotic regions marked by
a 2) were created by a trajectory within the pretzel region.}%
\label{fig:poincuneqRN}%
\end{figure}

This can be shown by creating Poincar\'{e} maps for the case when all three
masses are unequal. An example for both the R and N system is shown in
Figure~\ref{fig:poincuneqRN} where the mass ratio is $m_{1}$:$m_{2}$:$m_{3}$ =
1:5:10. Here we see that the symmetry about the $p_{R}=0$ axis no longer
exists in the Newtonian system due to the fact that none of the particles have
equal mass. We also see a further warping of the relativistic plots due to
this added asymmetry. Furthermore, we find that the different regions are not
as clearly segregated as in the $m_{1}=m_{2}$ phase space but extend over more
of the Poincar\'{e} map. For instance, in the relativistic map of
Figure~\ref{fig:poincuneqRN} we see that the chaotic region separating the
annulus and pretzel trajectories (marked by a 1) is no longer a single,
densely filled loop but actually two loops which were created by a single
trajectory. The annulus region is confined to the area inside both of these
loops, where a single annulus trajectory will visit both regions.

Besides this novel partitioning of the different regions, the changes to the
structure of the phase space for different ratios of the mass when all three
masses are unequal are analogous to the results described above for the case
where $m_{1}=m_{2}$. For example, the ratio 1:5:10 exhibits similar behaviour
as the case when there is one light particle and two heavy ones, only the
effects are not as prevalent because of the intermediate mass particle. The
results can be seen as an interpolation between the 1:1:1 case and the 1:10:10 case.

One major difference that we find between the equal and unequal mass cases is
the presence of additional chaotic regions in the unequal mass space that are
not present in the corresponding constant energy hyper-surface of the equal
mass space. This is true for both the Newtonian and relativistic systems. For
the mass ratios and energy levels that we have studied, these additional
chaotic regions appear within the pretzel regions of the corresponding equal
mass surface of section. The novel chaotic trajectories are characterized by
broadened lines in the pretzel region as can be seen in
Figures~\ref{fig:poinclowalphaRN}--\ref{fig:poincuneqRN}. The exact, physical
mechanism that gives rise to this increase in chaos is not presently known. \bigskip

Note that, although there are no apparent regions of chaos in the Newtonian,
$\alpha=0.1$ Poincar\'{e} map of Figure~\ref{fig:poinclowalphaRN} (besides the
one separating the annuli from the pretzels), we do find a slight broadening
of lines in the other two Poincar\'{e} maps constructed by plotting points
each time the hex particle crosses the other two bisectors respectively.

That is, additional regions of chaos do form in the unequal mass phase space
but these new regions cannot be seen on the particular choice of Poincar\'{e}
section shown in Figure~\ref{fig:poinclowalphaRN}. We suspect that these new
regions of chaos would become more prevalent as the difference in particle
masses increases.

It is instructive to compare our results with a similar study of a billiard in
$R^{2}$ colliding with a wedge in a uniform (Newtonian) gravitational field
performed by Lehtihet and Miller (referred to as LM herein) \cite{LMiller}. LM
showed that the two-dimensional wedge billiard system is isomorphic to a
system of 3 elastically colliding, self-gravitating particles (under Newtonian
gravity) in one-dimension, with the relative masses of the particles directly
related to the wedge angle by
\begin{equation}
\tan\theta=\frac{\sqrt{1+2\alpha^{-2}}}{1+2\alpha^{-1}} \label{massangle}%
\end{equation}
where $\alpha$ is as defined in our study. LM only considered the situations
where the wedge is symmetric, which corresponds to the case when two of the
three masses are the same. The value of $\theta=\pi/6$ corresponds to
$\alpha=1$, the equal mass case. This connection between particle masses and
the wedge angle agrees with the distortion of the potential energy described
in \S~\ref{sec:genprop} where the angle of the wedge is related to the angle
between the bisectors of the hexagonal well.

The only difference between the LM system and our N system is the existence of
collisions in the former while the particles pass through each other in the
latter. For the case where all three particles are identical, it is irrelevent
whether one considers that the particles are colliding or passing through each
other (besides the question of labelling the particles). For this reason, the
phase space structure of the Newtonian equal mass configuration, as presented
in the Poincar\'{e} maps, is identical between our system and the
wedge-billiard system.

LM found that this wedge-billiard system exhibits the characteristics of a
conservative Hamiltonian system with two degrees of freedom and a
discontinuity. By changing the value of a single continuous parameter,
$\theta$, they found a variety of dynamics similar to our study. More
specifically, for $\theta<\pi/4$ (which corresponds to the entire range of
physical values of $\alpha$) they found that integrable, near-integrable
(KAM), and chaotic regions coexisted in phase space. Furthermore, as the wedge
angle was increased from $\pi/6$ (corresponding to both an increase \emph{or}
a decrease of $\alpha$ due to the nature of the connection between mass ratio
and wedge angle (\ref{massangle})), they found that the region surrounding
periodic fixed points was consumed by regions of simply connected chaos which
increased in size with increasing wedge angle.

As noted above, we find a similar behaviour in both our Newtonian and
relativistic systems in that we see an increase in the amount of chaos as the
difference in the masses increases. However, we have only studied moderate
particle mass differences in order to characterize the general nature of the
unequal mass system and it is not clear how the global structure of our system
will behave for very large differences in the particle masses. In particular,
we do not know if our systems will experience a global transition to chaos or
if there exists integrable and near integrable regions for all mass ratios.
This remains an area for further study.

\section{Discussion}

\label{sec:disc} We have presented the results of a continued study of the
3-body problem in lineal gravity begun in \cite{3bdshort, burnell}. The focus
of the present investigation was to see what happens to the motion of the
particles when the relative masses of each are not equal. Here we summarize
our results.

The derivation of the 3-body Hamiltonian by canonical reduction of the action
(\ref{act1}) was summarized and the associated post-Newtonian and Newtonian
Hamiltonians presented. Each Hamiltonian possesses two spatial degrees of
freedom with two conjugate momentum degrees of freedom and these were made
manifest by changing to $(\rho, \lambda)$ coordinates. Expressions for the
potential energy of each system were derived and the distortion of the
potential energy due to varying the mass ratio was described.

The results of the study of the equal mass case were summarized and the
different types of motion were classified into three categories: annulus,
where each particle always crosses the other two in succession; pretzel, in
which two particles can cross each other twice in a row; and chaotic, where
the sequence of particles crossings does not progress in a discernible
pattern. By studying the motion of the 3 particles and their corresponding
hex-particle representation in the $\rho$-$\lambda$ plane, we characterized
how changing the mass ratio of the particles effects the dynamics of the
system. More specifically, we described in physical terms how the type of
motion (annulus, pretzel, and chaotic) and their relative abundance in phase
space changes with respect to the mass ratio.

As the relative difference between the masses of the particles increases, we
find the onset of additional regions in phase space of chaos that are not
present in the equal mass system \textbf{-- }in other words, we find that
motion that was once quasi-periodic is now chaotic. \ This shows that the
unequal mass phase space is not simply a deformation of the corresponding
equal mass space but, indeed, contains novel dynamics. The physical mechanism
behind this phenomenon is currently unknown.

This is similar to the behaviour of a billiard colliding with a wedge (which
is isomorphic to 3 particles elastically colliding on a line under their
mutual, Newtonian attraction) studied by Lehtihet and Miller \cite{LMiller}.
It is still not known what happens to these novel regions of chaos as the
difference in mass gets exceedingly large.

There are still many open areas of study in the lineal, 3-body problem. As was
done in the 2-body problem \cite{2bd, 2bdcossh, 2bdcoslo, 2bdchglo} it will be
interesting to see the effect of adding charge to the particles and a
cosmological constant to the system. Furthermore, more sophisticated numerical
techniques need to be introduced in order to probe the dynamics of the system
at high energies and in order to study the motion in the post-Newtonian system
for unequal mass particles. A description of the global structure of phase
space for extreme differences in the particle masses is still needed in order
to determine the stability of the system in these limiting cases. As mentioned
above, a discrete map between particle crossings in the N and R systems
(although it is doubtful whether this can be obtained for the latter) may
illuminate some of the more general features of the 3-body system. The
development of a relativistic 3-body system where the particles elastically
collide instead of passing through each other would also be an interesting
subject to study to see if the increased chaos reported in \cite{LMiller} has
an analogue in the relativistic system.

\section*{Acknowledgements}

This work was supported by the Natural Sciences and Engineering Research
Council of Canada.

\end{document}